\documentclass[12pt,preprint]{aastex}

\newcommand{\eg}{{\it e.g.}\ }

\newcommand{\ie}{{\it i.e.}\ }

\newcommand{\etal}{{\it et al.}\ }

\newcommand{\ps}{Pan-STARRS}

\def\azpress#1 #2 #3 #4 #5 #6 {{#1} #2. #3. In {\it #4} (#5, Eds.), pp.\
#6.  Univ. of Arizona Press, Tucson.}

\begin{document}

\bigskip

\bigskip

\title{Efficient intra- and inter-night linking \nl
 of asteroid detections using kd-trees} 

\bigskip

\bigskip

\author{Jeremy Kubica$^{1\dagger}$, Larry Denneau$^2$, Tommy Grav$^2$,
James Heasley$^2$, Robert Jedicke$^2$, Joseph Masiero$^2$, Andrea
Milani$^3$, Andrew Moore$^{1\dagger}$, David Tholen$^2$, Richard
J. Wainscoat$^2$}

\bigskip
\bigskip

\affil{$^1$The Robotics Institute, 5000 Forbes Avenue, 
  Pittsburgh, PA, 15213-3890}  

\affil{$^2$Institute for Astronomy, University of Hawaii, 
  Honolulu, HI, 96822}  

\affil{$^3$ Dipartimento di Matematica, Via Buonarroti 2, 56127 Pisa, Italy}  

\affil{$^\dagger$ now at Google, Collaborative Innovation Center, 4720
Forbes Avenue, Pittsburgh, PA, 15213}

\bigskip
\bigskip

\author {\today}

\bigskip
\bigskip

\begin{center}
\bigskip
\bigskip
{\bf Efficient intra- and inter-night linking \\
 of asteroid detections using kd-trees} 
\end{center}

\bigskip

\begin{center}
\end{center}

\bigskip
\bigskip
\bigskip
\bigskip

\noindent {27 manuscript pages\\14 figures\\7 tables}
\begin{description}
\item[Keywords:] Asteroids
\end{description}
\clearpage

\noindent{\bf Running Head:\\}
\noindent{Efficient linking of asteroid detections} \\

\bigskip
\noindent {\bf Editorial Correspondence and Proofs:\\}
\noindent Robert Jedicke\\
Institute for Astronomy\\
University of Hawaii\\
Honolulu, HI, 96822\\ 
\\Tel: 808.956.9841\\
Fax: 808.988.8972\\
E-mail: jedicke@ifa.hawaii.edu

\clearpage

\begin{abstract}

The Panoramic Survey Telescope And Rapid Response System (Pan-STARRS)
under development at the University of Hawaii's Institute for
Astronomy is creating the first fully automated end-to-end Moving
Object Processing System (MOPS) in the world.  It will be capable of
identifying detections of moving objects in our solar system and
linking those detections within and between nights, attributing those
detections to known objects, calculating initial and
differentially-corrected orbits for linked detections, precovering
detections when they exist, and orbit identification.  Here we
describe new kd-tree and variable-tree algorithms that allow fast,
efficient, scalable linking of intra and inter-night detections.
Using a pseudo-realistic simulation of the Pan-STARRS survey strategy
incorporating weather, astrometric accuracy and false detections we
have achieved nearly 100\% efficiency and accuracy for intra-night
linking and nearly 100\% efficiency for inter-night linking within a
lunation.  At realistic sky-plane densities for both real and false
detections the intra-night linking of detections into `tracks'
currently has an accuracy of 0.3\%.  Successful tests of the MOPS on
real source detections from the Spacewatch asteroid survey indicate
that the MOPS is capable of identifying asteroids in real data.

\end{abstract}

\clearpage

\section{Introduction}
\label{s.Introduction}

The next generation of wide-field sky surveys will be capable of
discovering as many solar system objects in one lunation as are
currently known.  Their unprecedented discovery rate coupled with
their deep limiting magnitudes will make targeted astrometric and
photometric followup observations impossible for the vast majority of
objects.  Thus, it is necessary that the new search programs employ
survey strategies that reacquire multiple observations of the same
objects within a lunation (a lunar synodic period).  Furthermore,
these facilities would tax the current capability of the International
Astronomical Union's Minor Planet Center (MPC), the clearing house for
observations of the solar system's small bodies, for linking the
detections and orbit determination.  The only solution is that the
surveys must provide the capability themselves and then provide the
MPC with pre-linked, vetted detections over multiple nights.
Simplistic linking algorithms for those detections scale like the
square of the sky-plane density ($\rho$) and, at the high densities
expected for the next generation surveys, the linking procedure could
dominate the processing time.  This work presents algorithms to solve
the problem that are fast, efficient, accurate and scale as $O(\rho
\log \rho)$.  We test our algorithms on pseudo-realistic simulations.

The history of asteroid orbit determination is mathematically rich.
It all began with the visual discovery of Ceres by Giuseppe Piazzi in
1801 and subsequent theory of orbit determination by \citet{Gau1809}.
At that time new techniques were developed to handle the orbit
determination from a short arc of observations and the ephemeris
errors on the observations of Ceres were many arcseconds.  Two hundred
years later absolute astrometric residuals are about an order of
magnitude better and the next generation surveys promise to reduce
those residuals for bright asteroids another order of magnitude.

As of 2006 August 6 there were a total of 338,470 asteroids in the
astorb database \citep{Bow94a} and over 20K asteroid observations are
reported daily to the Minor Planet Center.  As new observations of
previously known asteroids are identified their orbital elements are
automatically updated.  Furthermore, new observations of asteroids
that were unknown are linked together and their orbits are calculated
quickly and automatically by digital computers.

The discovery rate of asteroids and comets has climbed dramatically in
the past decade (for an overview of current asteroid search programs
see \citet{Sto02}) due to the advent of new technologies like the CCD
camera and because of NASA's Congressional mandate to search for Near
Earth Objects (NEO) larger than 1km in diameter \citep{Mor92}.  The
mandate to identify 90\% of NEOs in this size range will most likely
be achieved shortly after the 2008 deadline \citep{Jed03}.

Asteroids (and often comets) are usually identified by their apparent motion
against background stars in an image during the time between three or
more exposures separated in time by tens of minutes.  All existant
surveys have relied on the nearly linear motion of the objects on the
sky during the short time between exposures to distinguish between
real objects and random alignments of false detections (noise).  Some
historical and contemporary surveys identify or check their
observations of moving objects by eye.

As the discovery rate and the limiting magnitude of the surveys has
increased the sky-plane density of asteroids has increased and, with
it, the opportunity for false identifications and linkages.  This
explosion in the number of reported observations to the MPC has
generated a corresponding theoretical examination of the techniques
used in linking new observations and fitting orbits
\citep[\eg][]{Mil05,Gra05,Kri04,Kri02,Vir01,Kri92,Mar85}.  These
problems, as well as that of attribution (identifying observations
with known objects), orbit identification (realizing that multiple
instances of an object's orbit appear in a database), and precovering
observations (identifying earlier detections of an object in a
database), are described by Milani \etal in a series of articles
\citep{Mil01,Mil00,Mil99a,Mil99b}.

This work describes new algorithms, and the testing framework
developed to measure their efficiency and accuracy, for intra and
inter-night linking of asteroid detections.  The algorithms work well
in simulations of the performance of the next generation sky surveys.

\section{\ps}
\label{s.PanSTARRS}

Spurred by the 2001 decadal review \citep{McK01} a new generation of
all-sky surveys are expected to commence operations within the next
ten years.  These new surveys will take advantage of the latest
developments in optical designs (\eg to produce large, flat fields of
view) and CCD technology (\eg extremely fast readout) to survey the
sky faster and deeper than ever before.  

The first of the next generation surveys to image the sky will be the
Panoramic Survey Telescope And Rapid Response System
\citep[\ps,][]{Hod04} located in Hawaii.  \ps\ will be composed of
four 1.8m diameter telescopes each with its own 1.44 Gpix camera
(0.3\arcsec/pixel).  Images from each of the four cameras will be
combined together electronically.  The cameras will use an innovative
new CCD technology composed of Orthogonal Transfer Arrays
\citep[OTA,][]{Ton97} that allow charge to be moved on the CCD in both
the $x$ and $y$ directions in real time at $\sim$30 Hz to compensate
for image motion due to the atmosphere or any tracking problems.  In
effect, the system produces a tip-tilt corrective optics on-chip
rather than with the secondary {\it and} it is able to achieve
superior seeing over the entire $\sim$7 deg$^2$ field-of-view rather
than just within the small isoplanatic angle in the center of the
field.  A prototype system (PS1) located on the summit of Haleakala,
HI, saw first light in the summer of 2006 and will begin science
operations in the summer of 2007.

One of the primary scientific goals of the \ps\ survey is to identify
90\% of all potentially hazardous objects larger than 300m diameter
within its ten year operational lifetime.  In the process it will
identify about 10 million other solar system objects.  It is expected
to reach $R\sim 24$ at 5-sigma in 30sec exposures at which level the
sky-plane density of asteroids will be about 250/deg$^2$ on the
ecliptic.  This is also the predicted density of false 5-sigma
detections in the image.  Thus, the ratio of false:real detections at
5-sigma is equal to unity on the ecliptic and increases dramatically
off the ecliptic.  Given enough computing power and/or time it is, in
principle, possible to link individual detections together on separate
nights of observation.  {\it A priori} distributions of asteroid
velocities and accelerations at any sky location could be used to
intelligently link detections on separate nights and then fit orbits
to them to select those that represent observations of objects.  (Note
that we distinguish between a detection, which is a set of pixels on
an image with elevated signal relative to the background, and an
observation which is a detection associated with a real object.)  This
method has not yet been used in practice because of the combinatorics
of the problem as the limiting magnitude of the system is approached
and the number of real and false detections increases dramatically.
It is almost certain that this technique will require 4 nights on
which each object was detected in order to determine orbits with good
fidelity.

As mentioned above, the typical contemporary asteroid survey obtains
$\ge$3 observations of an asteroid within a short period of time on a
night.  When these observations are submitted to the MPC there is high
probability that each set of detections corresponds to a real object.
The MPC's responsibility is to link these detections to known objects
or to other new detections of the same object.  Many of the
contemporary and all the historical asteroid surveys identified NEOs
through their anomalous rates of motion relative to other objects in
or near the field of view \citep{Jed96}.  

There are two main problems with this mode of operation for the next
generation surveys.  First, in order to guarantee that reported sets
of detections correspond to real objects, surveys require $\ge$3
detections on a night which dramatically limits the system's sky
coverage; \eg a system that obtains only 2 detections/night can cover
50\% more sky, and obtain 50\% more detections than a survey requiring
3 detections/night.  Second, follow-up of NEO detections for the
contemporary surveys is typically accomplished by the survey itself or
by other professional surveying systems.  Since the first next
generation survey (at least) will not have the luxury of any other
existing system being able to recover newly discovered objects, the
survey must obtain its own follow-up.

The \ps\ system will most likely obtain just 2 images per night of
each solar system survey field but re-image the field 3 or 4 times
within a lunation.  Two images are used each night in order to
distinguish between false and real detections and separate stationary
and moving transient objects.  It also has the benefit of providing a
small motion vector for each possible observation.  Obtaining the same
object a few more times within the next two weeks provides both
recovery of the objects and more nights of observations with which to
calculate an orbit and verify the reality of each set of detections.
Since it is (currently) required that detections reported to the MPC
have a high probability of being legitimate observations, \ps\ will
only report those detections to the MPC that are linked across nights
into real orbits.  Thus, \ps\ must develop the capability of linking
detections across nights into real orbits.  If the MPC relaxes the
condition on the accuracy of linked detections then \ps\ will report
everything that is available.

The responsibility for intra-night (within a night) and inter-night
(between many nights) linking of detections (as well as attributing,
precovering, orbit determination and identificationm, etc.) rests with
\ps's Moving Object Processing System (MOPS).

\section{Pan-STARRS Moving Object Processing System (MOPS)}
\label{s.Pan-STARRS-MOPS}



Images from the cameras on each of the four \ps\ telescopes (for the
\ps-1 system only a single camera and telescope will be in operation)
are first passed through the Image Processing Pipeline (IPP) that
aligns, warps, removes cosmic rays, etc., and digitally combines them
into a single master image.  Many master images are combined together
to create a high S/N static-sky images that is subtracted from the
current master image to obtain a difference image containing only
transient sources (stationary and moving) and noise (false
detections).  The difference image is then searched for sources
consistent with being asteroids (both nearly stationary and moving
fast enough to trail) and also for comets.  Pairs of difference images
separated by a Transient Time Interval (TTI) of about 15-30 minutes
(the time separation is still to be determined and may vary with
sky-plane location) are analyzed in the same manner.  A list of all
the identified sources in both images along with their characteristics
(time, trail length, axis orientation, flux, etc.) is then passed to
the MOPS.  The software and algorithms described herein are expected
to be applicable to both \ps-4 and \ps-1 and the tests described
herein are performed at asteroid sky-plane densities (\ie limiting
magnitude) expected for the four telescope system.

The MOPS will;
\begin{itemize}
\item link intra-night detections into probable observations (tracklets),
\item attribute tracklets to known objects,
\item link inter-night detections into possible objects (tracks),
\item perform an initial orbit determination (IOD) to select tracks
that are likely to be real objects,
\item perform a differential correction to the orbit determination (OD)
to obtain a derived orbit for the track,
\item identify whether an earlier derived orbit is identical to the
current orbit,
\item seek precoveries in all earlier images of the derived object,
\item and determine its operational efficiency and accuracy in nearly 
real time using a synthetic solar system model.
\end{itemize}

The results described herein only describe the algorithms and
efficiency for the first and third steps, the intra-night and
inter-night linking of detections.  The performance of the MOPS for
the other aspects of its operation will be described in future papers.

At 5-sigma (or $r\sim 24$mag) we expect about 250 false
detections/deg$^2$ \citep{Kai04} or about 1750 false detections per
image at any position on the sky.  To the same S/N we also expect a
maximum sky-plane density of asteroids on the ecliptic of about
250/deg$^2$ \citep{Gla06,Mas06,Yos03} but this number decreases
dramatically off the ecliptic.  At 3-sigma the false detection rate
will be about 100$\times$ higher with only an increase of about
1.4$\times$ in the number of real detections.  It is clear from these
ratios that the difficulty of identifying asteroid observations
increases substantially as we push the limiting operational S/N into
the noise.  The S/N at which the \ps\ MOPS will operate will be
determined when the actual operational characteristics of the system
are known.  For this work we assume a 5-sigma cutoff corresponding to
$r \sim 24$mag for the four telescope \ps\ facility.

The first step in the MOPS is to identify sets of detections in images
within a night that are spatially close and therefore likely to be
observations of a real object.  We call these sets of detections {\it
tracklets}.  The MOPS also uses trailing information in the form of
the length and orientation of each detection to further constrain the
intra-night linking problem - only those detections that have the
expected trail length and orientation given their separation in time
and space are combined into tracklets.  We will demonstrate below that
our process is almost 100\% efficient at identifying tracklets with an
accuracy in the range of 85-90\% (see table
\ref{tab.StandardTrackletPerformance}).

The second MOPS step is the inter-night linking of tracklets into sets
that we call {\it tracks}.  In operations this step is followed by IOD
and OD to select only those tracks that are valid orbits.  We will
show below that at the expected sky-plane density of real and false
detections the set of realized tracks are mostly false.  But after IOD
and OD we are left with a nearly pure sample of actual orbits.  The
key is to use the track formation process to reduce the number of
false tracks to a sufficiently small number that it is feasible to
calculate orbits for all tracks within the required time frame.

The difficulty in intra- and inter-night linking of detections is
combinatoric and increases like $\rho^2$, where $\rho$ is the number
of detections/deg$^2$, if a brute-force approach is taken in linking
the detections.  A few sophisticated techniques have been proposed
\citep[\eg][]{Gra05,Mil05} to deal with these problems.  We report
here on our success with a linking algorithm that makes use of a
clever data structure (known as a kd-tree) to convert the combinatoric
problem in both cases into one that increases instead like $\rho \log
\rho$.  In this manner we can explore and reject many possible
linkages without resorting to sophisticated and time consuming orbit
determination techniques and thereby increase the speed with which we
can manage the large number of detections (false and real) from the
next generation surveys.

\section{Solar System Model}
\label{s.SolarSystemModel}

To verify that our linking algorithms are efficient we require a model
of the various populations of small bodies in our solar system that
could possibly reach $r\sim 24.5$.  This simulation requires realistic
orbits rather than simply the objects' spatial distribution.  These
requirements forced us into developing our own Solar System Model
(SSM) rather than adopting \citet{Ted05}'s Statistical Asteroid Model
(SAM) for main belt asteroids, though we were motivated by some of the
techniques developed for the SAM.

Our SSM will be discussed in detail by \citet{Den06a} and only briefly
here \citep[also][]{Mil06}.  For the purpose of testing the MOPS we
have developed a preliminary model of many populations of objects in
our solar system and beyond including nearly 11 million small bodies:
\begin{itemize}
\item Near Earth Objects (NEO) (including objects entirely interior to the
Earth's orbit)
\item Main Belt Objects (MBO)
\item Jupiter trojans and trojans of all other planets
\item Centaurs (CEN)
\item Jupiter Family, Halley-type and Oort Cloud comets (COM)
\item Trans-Neptunian objects (TNO) - classical, resonant, scattered
and extended scattered disk.
\end{itemize}

The details of the model are not critical to interpreting the work
reported here.  In general, we have a preliminary model of different
small body populations in the solar system (and some populations that
have not been discovered) that mimic the real objects at different
levels of fidelity in each of the following properties:
\begin{itemize}
\item orbit distribution
\item absolute magnitude (H), size and albedo distribution
\item shapes modelled as tri-axial ellipsoids
\item rotation rates
\item pole orientations
\end{itemize}
For the simulations described here we simply used the absolute
magnitude and standard formulae \citep{Bow89} for converting to
apparent magnitudes rather than incorporating the shape, rotation rate
and pole orientation.

The input orbit distributions for the NEOs \citep{Bot02} and CENs
\citep{Jed97} have a pedigree traceable to published studies while the
MBOs mimic the large statistics of the nearly complete MBO population
(for $H<14.5$) \citep{Jed02}.  For the moment, the input orbit
distributions for the other populations are based only on the observed
rather than the debiased populations.  In all cases we generated a
full suite of objects that might achieve $r<24.5$ (the expected \ps-4
limiting magnitude) at some time in the next ten years.

The absolute magnitude distributions were generated according to
corrected $H$ distributions where available (NEO - \citet{Bot02}, MBO
- \citet{Jed02}, CEN - \citet{Jed97}, TRO - \citet{Jew00}, TNO -
\citet{Ber04}, SDO - \citet{Ell05}).  For all types of comets the
absolute magnitude distribution was simply the observed distribution
extended to smaller sizes in a natural manner.  It is our intention to
improve this model for comets in the final solar system model
implement for MOPS.

\section{Survey Simulation}
\label{s.SurveySimulation}

Many researchers have modelled asteroid surveys in an attempt to
predict the performance of a particular system
\citep[\eg][]{Ray04,Mig02}.  Others have modelled generic survey
systems in order to elucidate more general principles
\citep[\eg][]{Jed03,Har98}.  For instance, in the case of discovering
NEOs, \cite{Bow94b} and \citet{Har98} showed that it is more important
to cover more sky than it is to go to fainter limiting magnitudes in a
smaller area.  These earlier simulations had a wide range of fidelity
to realism with some merely postulating that the entire sky would be
covered in a night.

The final mode of solar system surveying for \ps\ will be under study
until regular asteroid surveying begins in earnest.  Even then, we
believe that a regular review of the survey strategy will be necessary
in an attempt to maximize the system efficiency.  The simulation of
the survey implemented here is our first-order vision that
incorporates many of the most important aspects of an efficient and
realistic survey that has as its highest priority the identification
of sub-km Potentially Hazardous Objects (PHO).  A full discussion of the
survey simulation and its impact on the MOPS asteroid discovery rates
is in preparation \citep{Den06b}.

With PHOs in mind we place a high emphasis on covering the
`sweet-spots', the sky at small solar elongation and small ecliptic
latitude where the sky-plane density of PHOs at \ps's limiting
magnitude is expected to be highest \citep{Che04}.  We take advantage
of the fact that asteroids tend to be brighter near the anti-solar
point and attempt to identify high inclination or nearby objects
surveying a wide area in both longitude and latitude near opposition.

\begin{figure}
\epsscale{1.1}
\plotone{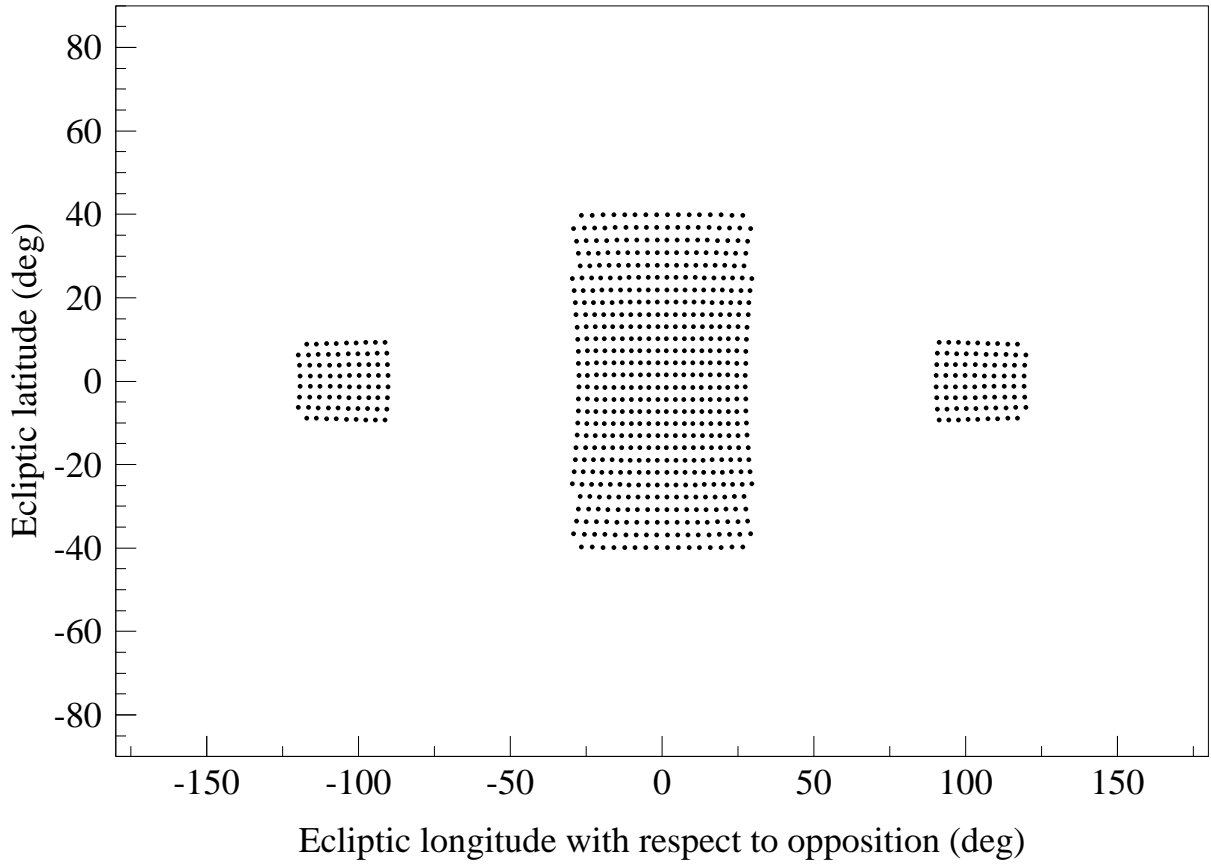}
\caption{828 equally spaced (in area) points in the
  $(\lambda^\prime,\beta)$ plane.  There are 660 points in the large
  opposition region in the center of the figure.  There are 84 points
  in each of the smaller sweet spot regions.  The evening(morning)
  sweet spot is on the left(right).\label{fig.hammer-aitoff.sss}}
\end{figure}

For the purpose of this work consider an ecliptic longitude
($\lambda^\prime$, opposition longitude) and latitude ($\beta$) system
centered on the opposition point. \eg opposition is always at $(0,0)$.
In this reference frame the solar system survey is defined by the two
sweet spots with $|\beta|<10\arcdeg$,
$-120\arcdeg<\lambda^\prime<-90\arcdeg$ or
$+90\arcdeg<\lambda^\prime<+120\arcdeg$ and also the opposition region
with $|\lambda^\prime|<30\arcdeg$ and $|\beta|<40\arcdeg$ totaling
about 5500 deg$^2$.  To simplify our simulation we assumed that the
\ps\ fields are square and of an area about equal to the final
expected camera field.  Figure \ref{fig.hammer-aitoff.sss} shows the
distribution of equal area field centers on the sky in the sweet-spots
and opposition regions.  There are 660 fields in the opposition region
and 84 in each of the sweet-spots corresponding to field coverage of
about 4,356 deg$^2$ and 1108 deg$^2$ (in both sweet-spots)
respectively.  Each \ps\ field covers about 7 deg$^2$ so this
simulation allows for some moderate overlap between adjacent
fields.

It is important to note that this scanning pattern (and the one likely
to be adopted for Pan-STARRS solar system survey operations) avoids
the (Main Belt) 'stationary spots' about 3.5 hours ($\sim50\arcdeg$) from
opposition.  The stationary spots are regions where apparent asteroid
motion along the ecliptic may briefly drop to zero.  The more distant
the asteroid population the greater the distance from opposition at
which the objects become 'stationary'.  For intance, TNOs are
stationary fully 80$\arcdeg$ from opposition - nearly in what we refer
to as the sweet-spots.  Asteroid paths on the sky can even form closed
loops far from opposition that might cause difficulty for the linking
algorithm desribed herein.

Moving objects will drift out of any fixed region on the sky.
Even a fixed-size region that moves at the mean rate of motion of
moving objects in the field will lose objects near its edge.  One
solution is to expand the size of the region with time.  Another
solution is to ensure that the region translates at a rate equal to
the mean rate of motion of the objects of primary interest in the
region.

We have used our solar system model (\S\ref{s.SolarSystemModel}) to
determine the apparent rate of motion of NEOs with $r<24$mag in the three
survey regions.  The sweet spots are small enough in ecliptic
longitude extent (30$\arcdeg$) that we included all NEOs in those
regions and found that they are moving at mean rates of
$d\beta/dt=0\arcdeg$/day (as expected from symmetry) and prograde at
$d\lambda\prime/dt\sim +0.65\arcdeg$/day.  The opposition region covers a
much wider range in ecliptic longitude and we are only in danger of
losing objects that are near its eastern and western edges.  Thus,
only those NEOs within 15$\arcdeg$ of the eastern or western edge of
the region were used to determine that they are moving retrograde at a
mean rate of $d\lambda\prime/dt\sim -0.30\arcdeg$/day.

For the purpose of this work we have assumed that the solar system
survey requires imaging of each field within a region three (3) times
per lunation with a mininum spacing of four (4) nights between any
successive visit to each field.  While this scenario is suitable for
this simulation we have evidence that another night of observation,
especially in the sweet-spots, will be necessary to resolve degenerate
multiple orbit solutions.  When running the simulation we have assumed
that a random 25\% of nights are entirely clouded out while the
remaining 75\% are entirely clear.  This results in a variable number
of nights between visits in a lunation.

The algorithm for scheduling the fields within the regions is
described below.  For the purpose of developing the inter-night field
scheduler it was convenient to think in terms of scheduling nights
with respect to full moon ($FM+N$ = Full Moon plus $N$ nights).
Evening and morning sweet spots may be acquired on the same night but
the opposition region was impossible to schedule in its entirety on a
single night.  We divided the opposition region into northern and
southern ecliptic latitudes that need to be acquired on separate
nights and may not be imaged on nights on which a sweet spot is
acquired (sweet spots also have higher priority).

\subsection{Evening sweet spot}
\label{ss.EveningSweetSpot}

Objects in the evening sweet spot are being overtaken by the Sun.  The
{\it first} opportunity to visit the ESS is just after full moon
($FM+4$ days), when the waning moon is no longer in the bright sky
after astronomical twilight ends.  The {\it last} opportunity to catch
the ESS is a few days after new moon ($FM+18$ days) before the young
moon enters the evening sweet spot.  

When scheduling surveying in the ESS it is impossible (due to weather
or the other \ps\ science survey requirements) to predict what night
will be the actual last night of observation.  Thus, on the first
possible night of surveying in the ESS we assume the worst case
scenario that the last possible night will be the last opportunity to
survey the same region at $FM+18$ days.  The last night then defines
the ESS region and we then work backwards from that location at a rate
of $d\lambda/dt=+0.65\arcdeg$/day to determine the location of the ESS
on any of the previous nights on which it is actually acquired.

\subsection{Opposition}
\label{ss.Opposition}

Scheduling of the opposition regions is constrained by
the moon appearing in those regions when it is full.  For both regions
we have assumed that the first day it is possible to acquire these
regions is at $FM+7$ days and the last is at $FM+21$ days.  

When scheduling the opposition regions we assume that the second night
will be acquired at new moon and define the actual field locations on
a specific night by translating the region at a rate of
$d\lambda/dt=-0.3\arcdeg$/day.

\subsection{Morning sweet spot}
\label{ss.MorningSweetSpot}

Objects in the morning sweet spot are also heading towards the Sun but
they have many months until they pass behind it because the Sun is
moving away from them faster than they approach it.  Thus, the
location of NEOs in the area of the MSS move away from the horizon
with time and the sky-plane location of NEOs improves with time as a
lunation progresses.  Surveying in the MSS may start just before new moon
($FM+10$ days) and is possible until the just-before-full moon enters
the morning sky ($FM+24$).

For scheduling the MSS region we simply survey the optimal MSS region
on the first possible day that it can actually be surveyed and
translate the region by $d\lambda/dt=+0.65\arcdeg$/day to determine
the location of the MSS on subsequent nights on which it is acquired.

\subsection{Nightly scheduling of fields}

Once the fields for a specific night have been selected they need to
be scheduled for that night taking into account a wide range of system
parameters and other factors.  \ps\ will eventually employ a dynamic
telescope scheduler that takes into account hundreds of relevant
factors.  While the \ps\ telescope scheduler is being developed, for
testing purposes the MOPS has adopted TAO (Tools for Automated
Observing, Paulo Holvorcem,
http://pan-starrs.ifa.hawaii.edu/project/MOPS/TAO/html/readme.html).
TAO is a macro-scheduler and as such it attempts to schedule all
fields on a single night as efficiently as possible.  There are far
too many TAO configuration parameters to discuss each in detail here.
Several important configuration parameters are:
\begin{itemize}
\item {\bf Number of images of each target} = 2
\item {\bf sky-plane location} \\ 
  Preferring low air mass due to poorer seeing, higher extinction and
  increased sky background at lower elevations..
\item {\bf field priority}
\item {\bf intra-night cadence requirements}\\ 
  15min between visits to the same field on each night.  The standard
  time between exposures on the same night is known as a Transient
  Time Interval or TTI.  There is a 50\% tolerance on the actual
  scheduling.
\item {\bf inter-night cadence requirements}\\ 
  No less than 3 nights between visits in a lunation.
\item {\bf exposure time} = 30sec
\item {\bf read out time} = 5sec
\item {\bf telescope slew rate} = 5$\arcdeg$/sec
\item {\bf time of night} = 5$\arcdeg$/sec
\item {\bf azimuthally dependent altitude limits} = 20$\arcdeg$ ($\sim$2.85 airmasses)
\item {\bf cloud cover}
\item {\bf seeing conditions}
\item {\bf Moon avoidance angle at full moon} =  45$\arcdeg$ (scales with phase)
\item {\bf Min/Max Sun Altitude} = -15$\arcdeg$\\
  Intermediate between nautical and astronomical twilight.
\end{itemize}

We ran the scheduler for ten years of synthetic surveying.  The
scheduling efficiency for the solar system fields is essentially 100\%
for those fields that are well above the minimum altitude (some of the
most southern opposition fields are always below the altitude limit
and some of the sweet spot fields may also be unavailable at certain
times of the year).  Due to `weather' some of the regions were not
covered 3 times in a lunation.  

\begin{figure}
\epsscale{1.0}
\plotone{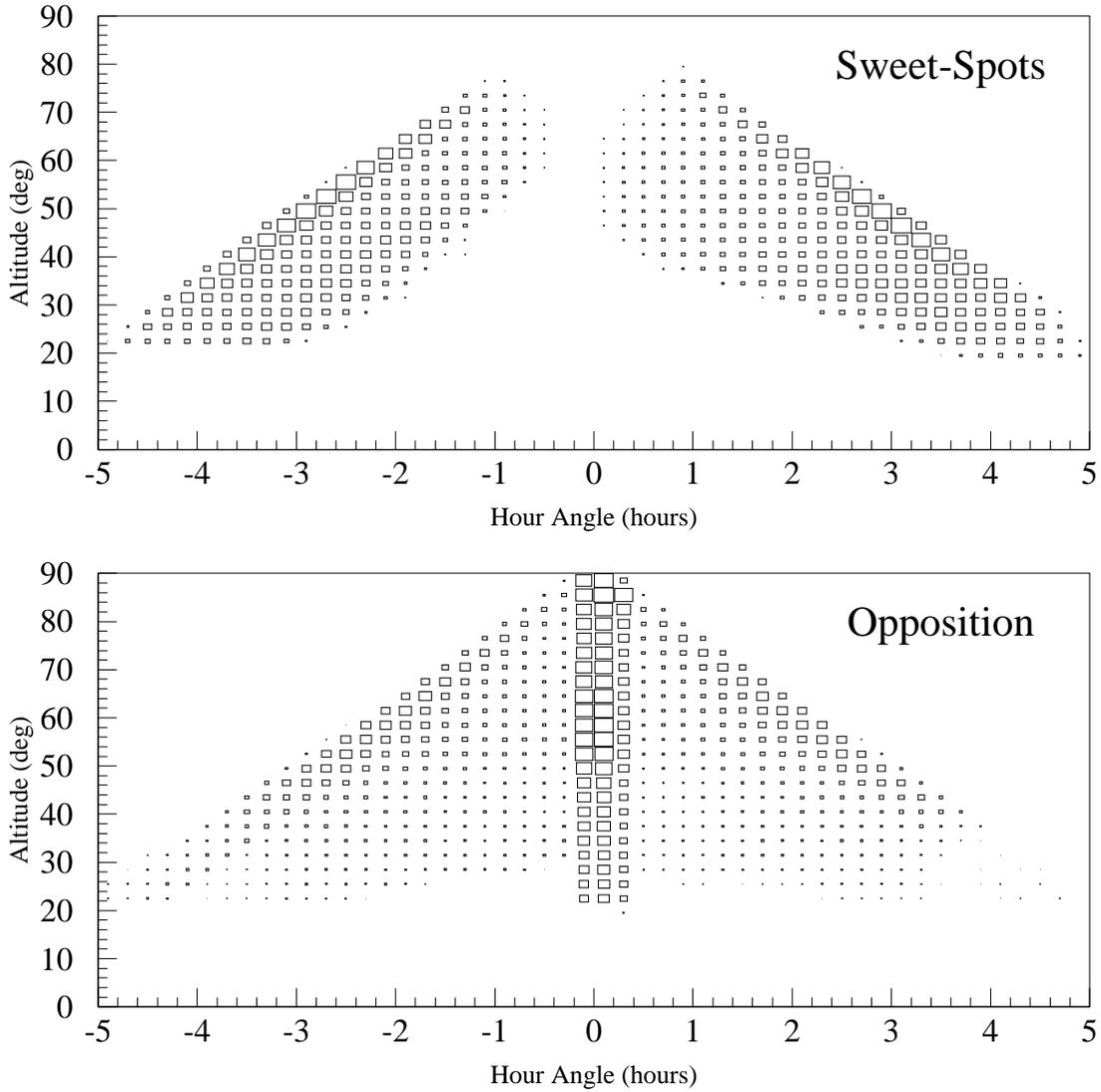}
\caption{Locations of field centers in altitude and hour angle for the
  sweet-spots (top) and opposition regions (bottom) in a ten year
  synthetic survey.  The size of a box is proportional to the number
  of fields acquired at that sky location.  Most of the opposition fields are
  acquired when they are at their optimal (highest) altitude near zero
  hour angles.  Most of the sweet-spot fields are acquired at the
  highest possible elevation for their hour angle.}
\label{fig.altaz}
\end{figure}

Figure \ref{fig.altaz} shows the distribution of field locations in
altitude vs. azimuth separately for the sweet spots and opposition
regions over the ten year survey.  The sweet spots are typically
obtained between 20$\arcdeg$ and 70$\arcdeg$ altitude and $60\arcdeg
<$ $|$azimuth$|$ $< 160\arcdeg$.  The most likely altitude is near
40$\arcdeg$ or about 1.7 airmasses.  For the opposition regions note
the predominance of fields scheduled near $\pm 180\arcdeg$ and close
to $0\arcdeg$ - on or near the meridian when the fields are at their
highest possible altitude (lowest possible airmass).

\section{Simulating detections}
\label{s.SimulatingDetections}

Given the survey simulation (\S\ref{s.SurveySimulation}) we generate
accurate n-body ephemerides and photometry for the synthetic solar
system objects (\S\ref{s.SolarSystemModel}) that appear in each field
of view.  The astrometric and photometric accuracy expected by \ps\ is
better than existing asteroid surveys.  At $r\sim 24$mag we expect
astrometric error to be about 0.1\arcsec\ and a photometric error of
about 0.35mag.  For brighter objects these errors will be considerably
smaller.

The linking method described herein is independent of the detection's
apparent magnitude except for the requirement that the detection be
above the limiting magnitude of the system (to simulate the expected
sky-plane density of asteroids).  However, in the interest of
completeness, and since modern orbit determination software can utilize
an estimate of the $S/N$, we generate a pseudo-realistic magnitude and
$S/N$ for synthetic detections.

The signal from a source of total apparent magnitude $m$ in an
exposure of time $t$ seconds and assuming PSF fitting photometry is
$S = t \times 10^{-0.4(m-m_1) / 2}$.  Assuming that the exposure is
sky-background limited, the variance from the sky is given by
$\sigma^2 = \pi \times FWHM^2 \times 10^{-0.4(\mu-m_1)} / 4$ where
$FWHM$ is the FWHM of the PSF in arc seconds and $\mu$ is the sky
brightness in magnitudes per square arcsecond.  The signal-to-noise
($S/N$) at magnitude $m$ is than given by
\begin{equation}
S/N =  PSN \times 10^{-(2m-M^\prime)/5} 
 \sqrt{ {1 \over \pi}  \biggl[{t \over sec }\biggr] \biggl[ {FWHM \over 1\arcsec} \biggr]^{-2} }
\end{equation}
where $PSN=$ 1 for PS1 and 2 for \ps\ while $M^\prime= \mu - m_1$
($\sim$45.6 for a $r$ filter in these simulations).  All the
simulations described herein involve the more difficult problem of
linking \ps\ rather than PS1 detections.

The astrometric error is assumed to be a symmetric 2-d Gaussian with
width given by:
\begin{equation}
\sigma = 0.01\arcsec + 0.070\arcsec \bigg[ { FWHM \over 0.6\arcsec } \bigg] 
  \bigg[ { 5 \over S/N } \bigg]
\end{equation}
In median seeing (with OTA correction in operation) and at $r\sim
24$mag we eventually expect an absolute astrometric accuracy of $\sim
0.07\arcsec$ with a minimum of about 0.01$\arcsec$ for bright,
unsaturated detections.

In order to automatically identify as many objects as possible the
MOPS will have to work in the presence of a substantial number of
false detections.  \citet{Kai04} estimates that at 5-$\sigma$ there
will be roughly 250 false detections/deg$^2$ - roughly the {\it
  maximum} number of actual objects in the same area.  To simulate the
presence of false detections in each image we generated random
locations in each field for each detection with a number density per
deg$^2$ given by:
\begin{equation}
\rho = 1.34 \bullet 10^7 \times  S/N * \exp\bigg[ - { (S/N)^2 \over 2 } \bigg]
\end{equation}

Most of the populations of objects in this work
(\S\ref{s.SolarSystemModel}) are moving slowly when detected in the
opposition and sweet spot regions but the NEOs may be moving fast
enough to leave small trails on the images.  We simulate this effect
by determining each object's rate and direction of motion and using
this information to determine the length and position angle of the
synthetic trail.

It is important to note some of the effects that we are not taking
into account in this simulation.  We believe that these effects are
not important in quantifying the efficiency of linking intra- and
inter-night detections.  By definition, an algorithm can only be
efficient at linking those detections that were identified.  So this
simulation implements a hard cutoff at $r=24$mag with 100\%
detection efficiency to that magnitude limit.  We do not account for
the camera CCD fill factor of $\sim$86\%, the fact that
almost 5\% of OTA `cells' on each camera will be used for image
guiding and lost to detecting moving objects, or a pre-processing step
implemented by Air Force space surveillance that will remove a few
percent of image pixels.  The fraction of pixels removed in the last
step will be a function of the time of night and sky-plane location
since more satellites will be visible towards sunset and sunrise than
at midnight.  We also do not account for astrometric and photometric
effects as a function of air mass.  \eg reduced astrometric and
photometric accuracy.

\begin{figure}
\epsscale{1.0}
\includegraphics[angle=270,scale=0.70]{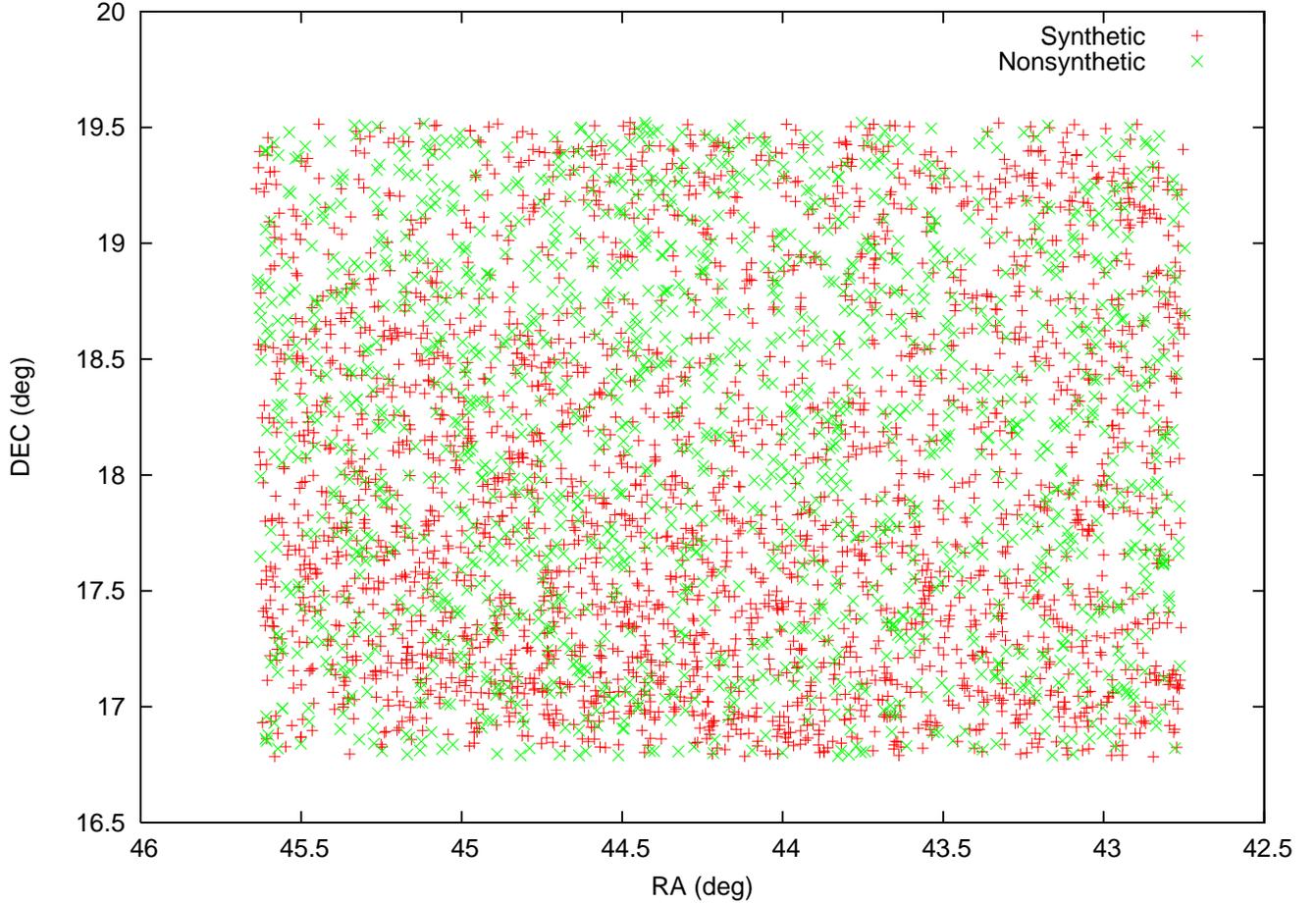}
\caption{A single \ps\ near-ecliptic field at full (\ps-4) density for both
  the solar system model (red $+$ symbols) and false detections (green
  $\times$ symbols).  The density of detections is about 250/deg$^2$
  on the ecliptic for each type.  The final \ps\ field will be in the
  shape of square chessboard with the four corner spots removed.}
\label{fig.field.detections}
\end{figure}

Figure \ref{fig.field.detections} shows a single field of synthetic
\ps\ detections.

\section{Linking detections}
\label{ss.LinkingDetections}

The preceding sections have outlined the input to the MOPS - a set of
transient detections of which a large fraction are false.  It is the
MOPS's responsibility to identify those detections corresponding to
observations of real objects.  The first step in this process is
identifying sets of detections that are nearby to each other spatially
and temporally and for which the distance between sequential
detections is consistent with an object moving at fixed speed.  We
call these sets of detections `tracklets'.  The second step is to link
tracklets together on multiple nights into `tracks'.  The brute force
approach to each of these steps would lead to prohibitively
CPU-intensive processing.  Instead, we have developed new techniques
using kd-trees to handle both these problems.  In the following three
sub-sections we introduce the concept of kd-trees and explain how
those data structures were applied to the MOPS requirements for intra-
and inter-night linking of detections.

\subsection{kd-trees}
\label{ss.kd-trees}

kd-trees are hierarchical data structures that can be used to
efficiently answer a variety of spatial queries \citep{Ben75}.
A kd-tree recursively partitions both the set of data points and the
corresponding space into progressively finer subsets and subregions.
Each node in the tree represents a region of the entire space and
(either explicitly or implicitly) a set of data points.

\begin{figure}[!tb]
\begin{center}
\begin{tabular}{ccccc}
\includegraphics[width=1.0in,angle=90]{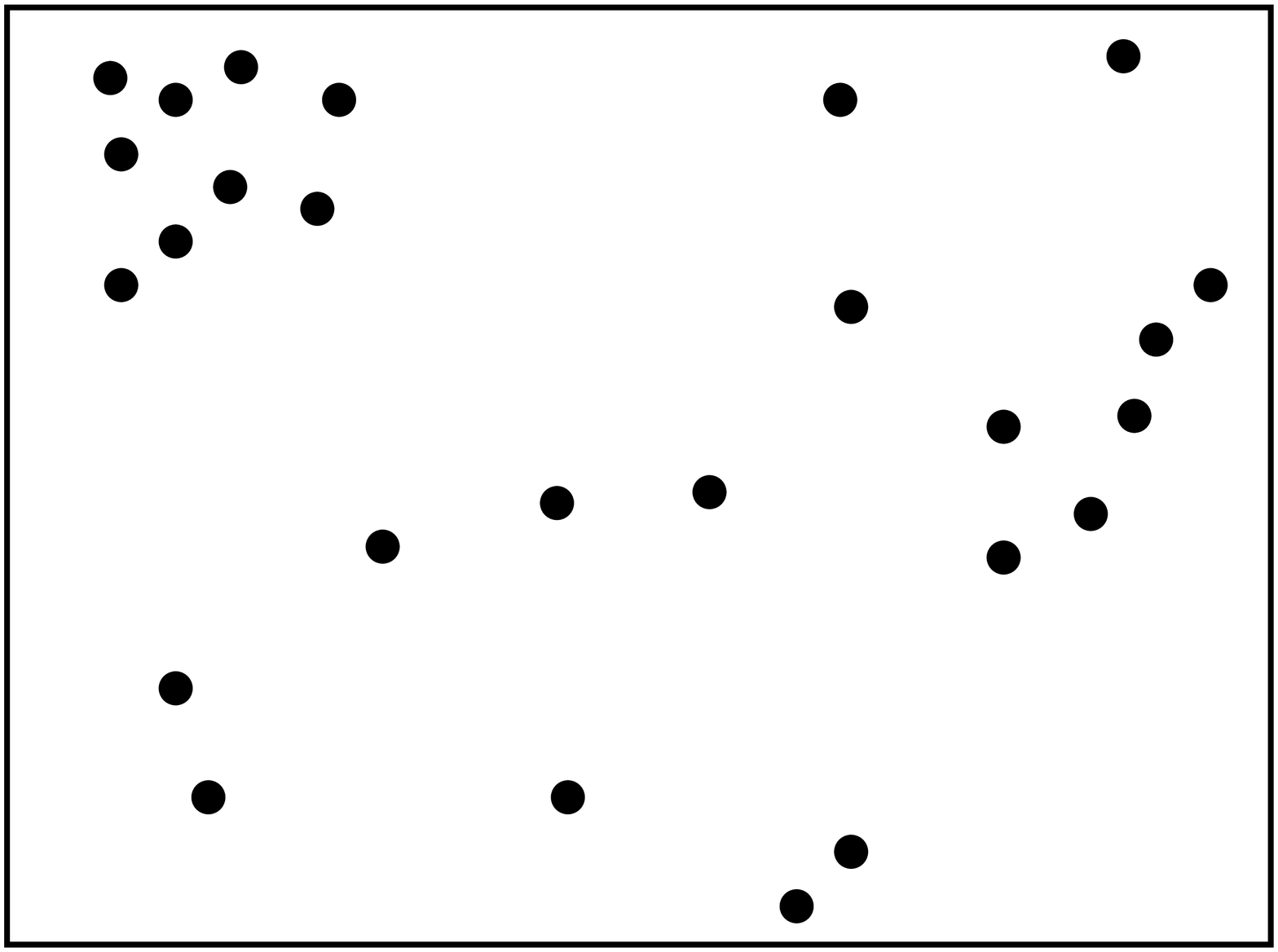} & \includegraphics[width=1.0in,angle=90]{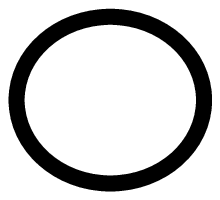} & & \includegraphics[width=1.0in,angle=90]{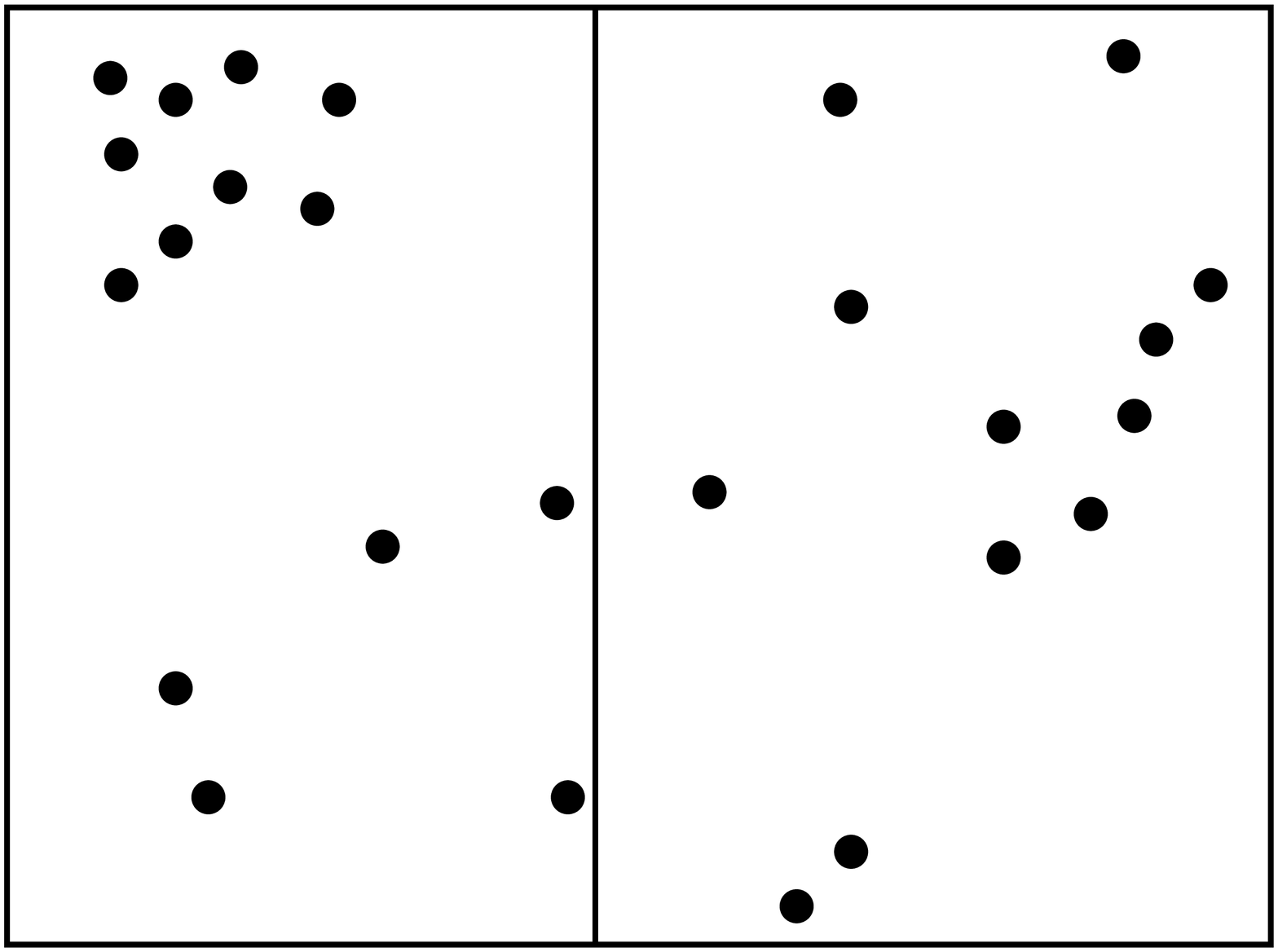} & \includegraphics[width=1.0in,angle=90]{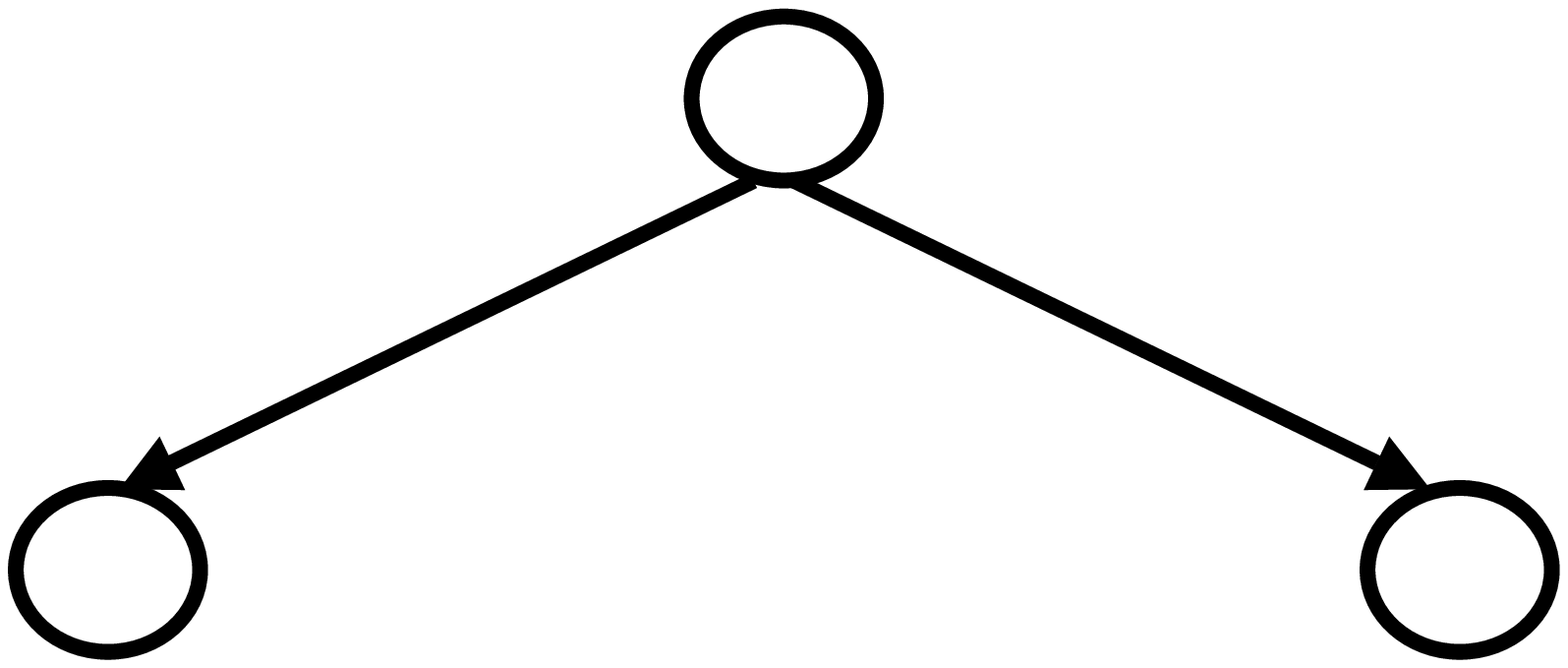} \\
\multicolumn{2}{c}{(A)} & & \multicolumn{2}{c}{(B)} \\
& & & & \\
\includegraphics[width=1.0in,angle=90]{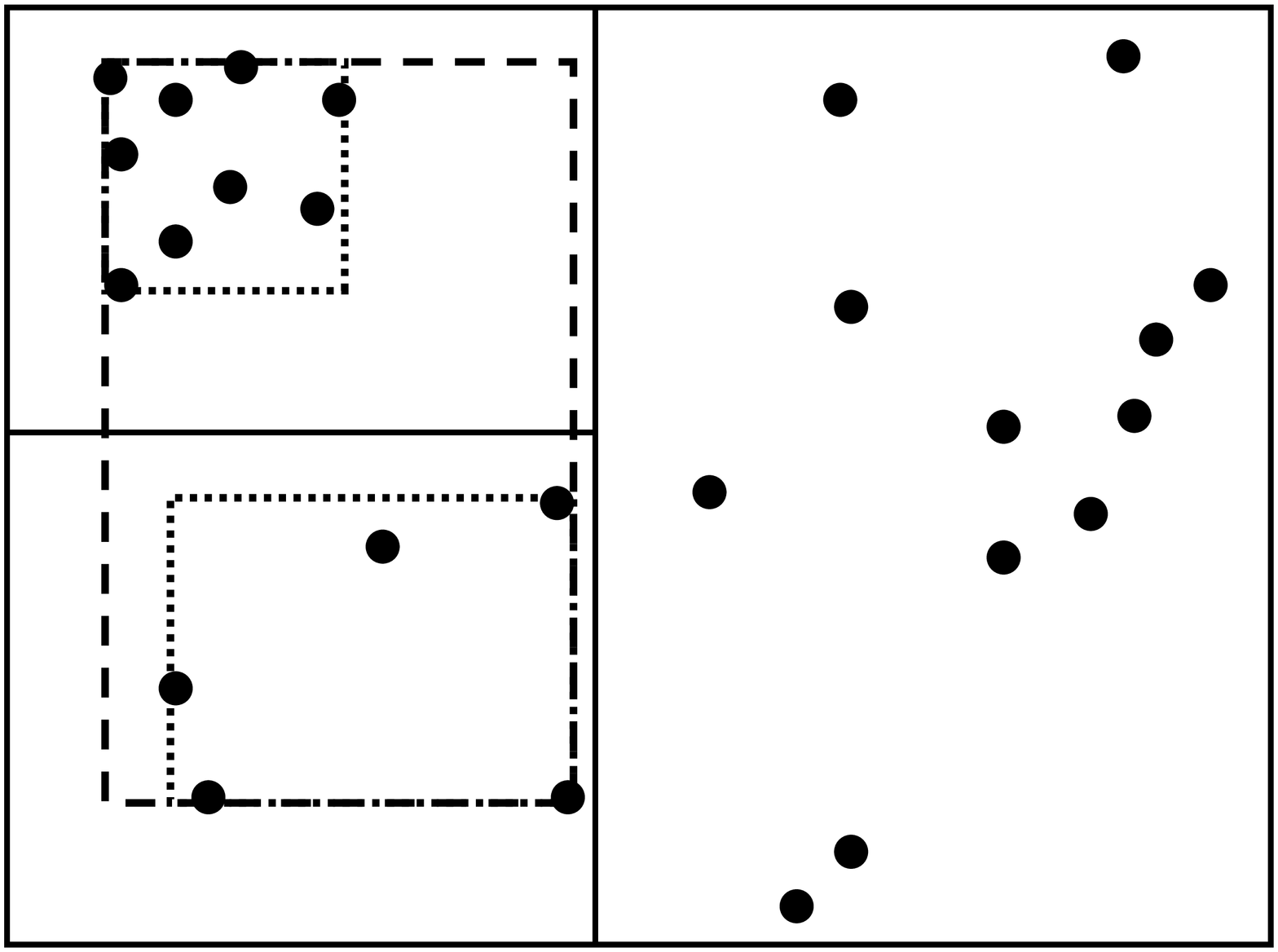} & \includegraphics[width=1.0in,angle=90]{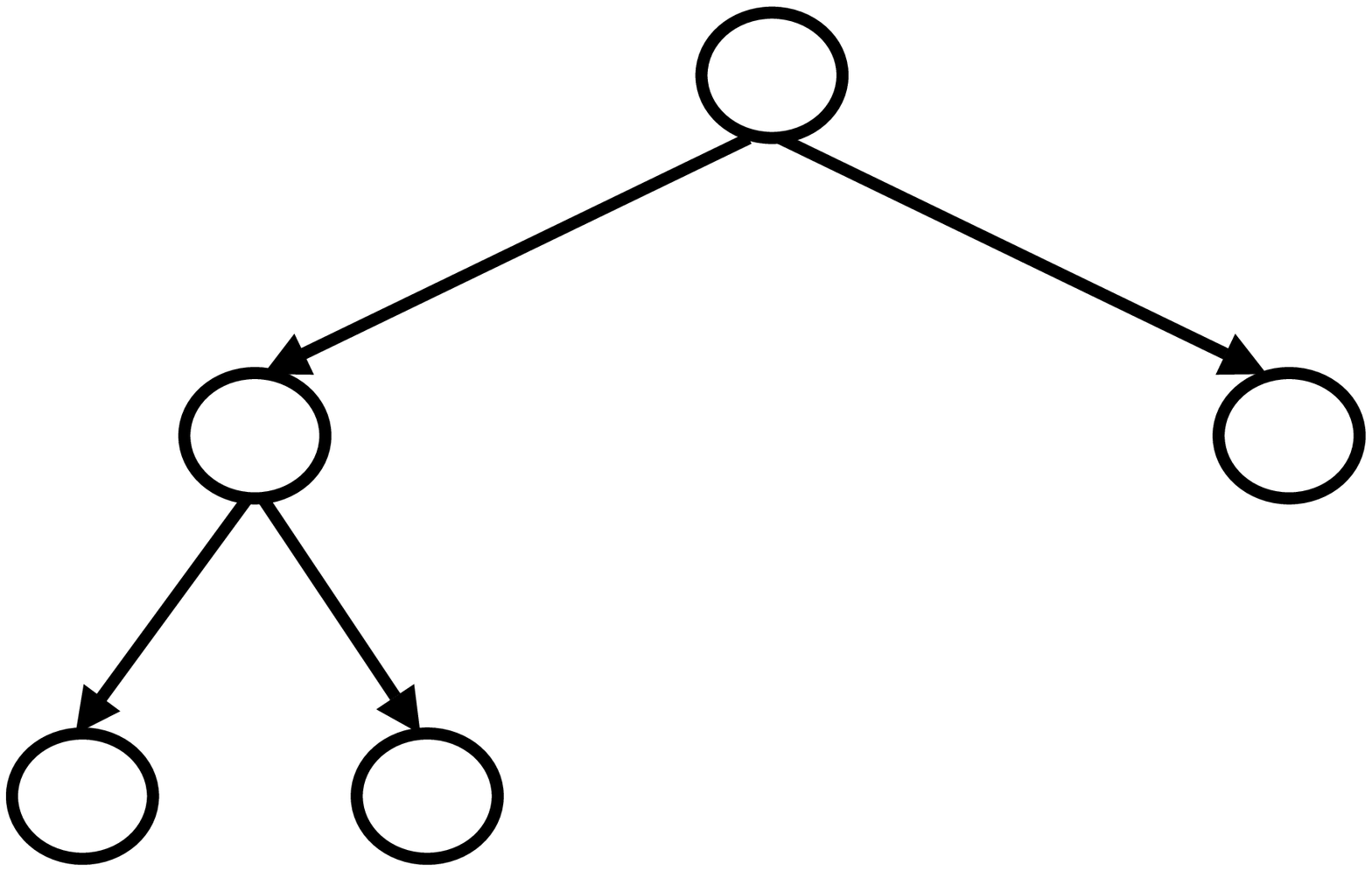} & & \includegraphics[width=1.0in,angle=90]{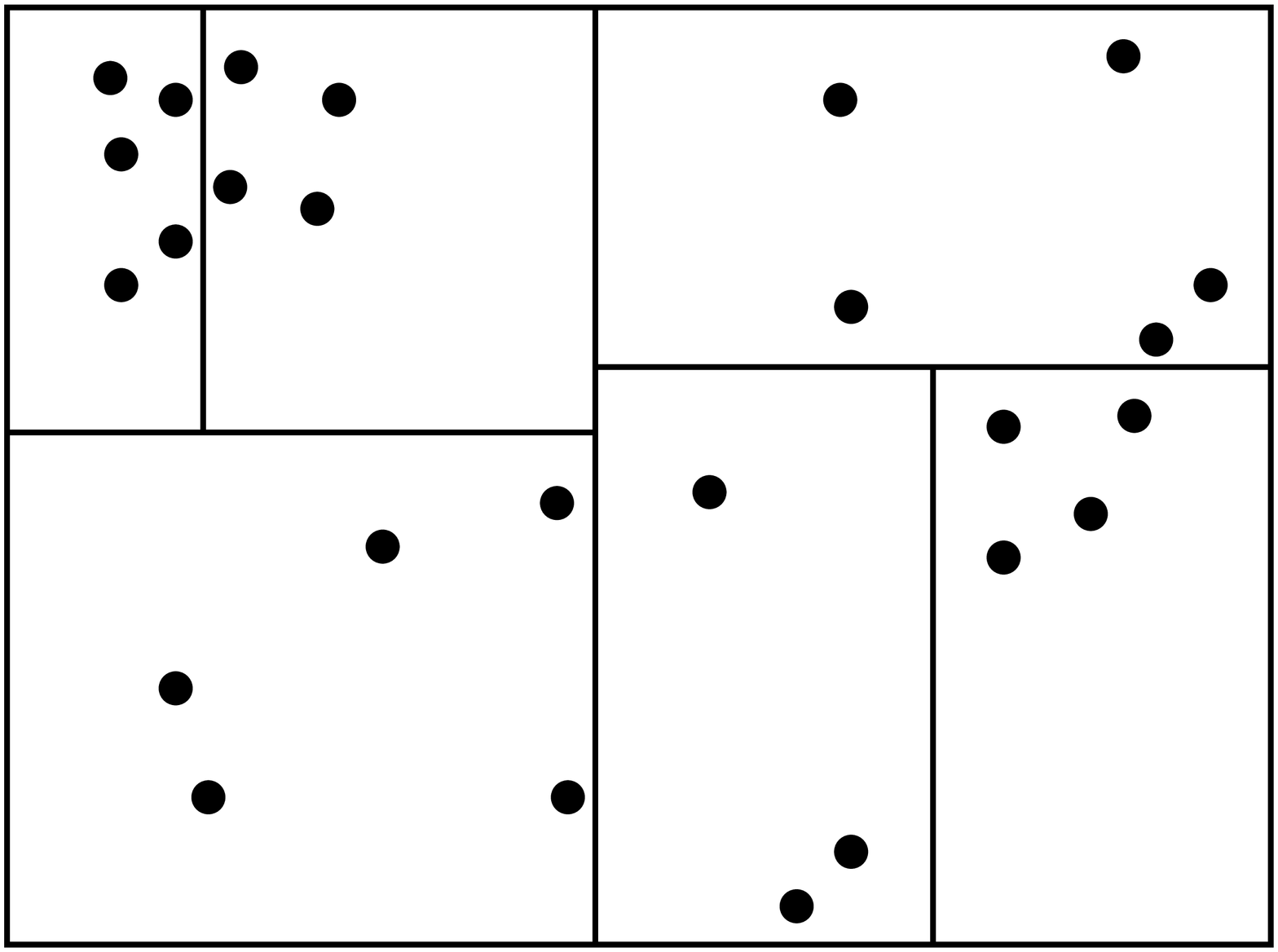} & \includegraphics[width=1.0in,angle=90]{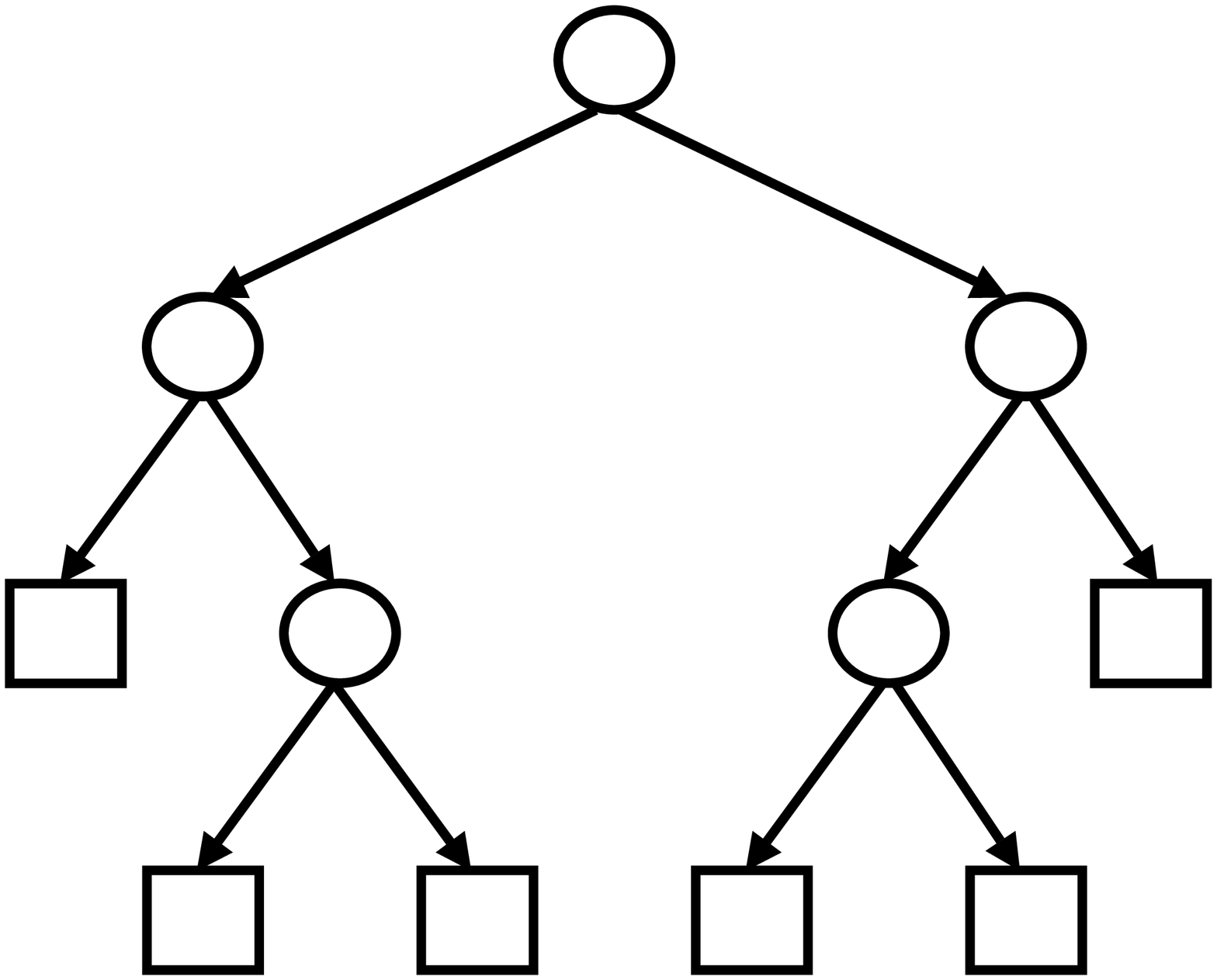} \\
\multicolumn{2}{c}{(C)} & & \multicolumn{2}{c}{(D)} \\
\end{tabular}
\caption{A kd-tree is constructed recursively in top-down fashion.  We
start with a single node containing all of the points (A).  This node
is split (B and C) into two subtrees.  At each level we calculate the
bounding box of the data owned by that node and store it in the node
(C).  C shows the bounding boxes for a node (dashed) and its two
children (dotted).  In the last figure (D) the final tree structure is
shown indicating that it is not necessary to have each leaf of the
tree contain only a single data point.}
\label{fig:kdbuild}
\end{center}
\end{figure}

A kd-tree is created in a top-down fashion as shown in
Figure~\ref{fig:kdbuild}.  At each level the current data is used to
calculate a bounding box for that node.  These bounds are saved and
stored at that node.  The data points are then partitioned into two
disjoint sets by splitting the data at the midpoint of the node's
widest dimension.  Each of these two sets is then used to recursively
create \emph{children} nodes.  We halt this process when the current
node owns fewer than a pre-established minimum number of points and
mark this node a \emph{leaf node}.  By the hierarchal structure of the
tree, the set of data points owned by a non-leaf node is the union of
its childrens' data points.  Thus we only need to explicitly store
pointers to the individual data points at the leaf nodes.

\begin{figure}
\begin{center}
\begin{tabular}{ccc}
\includegraphics[width=1.25in,angle=90]{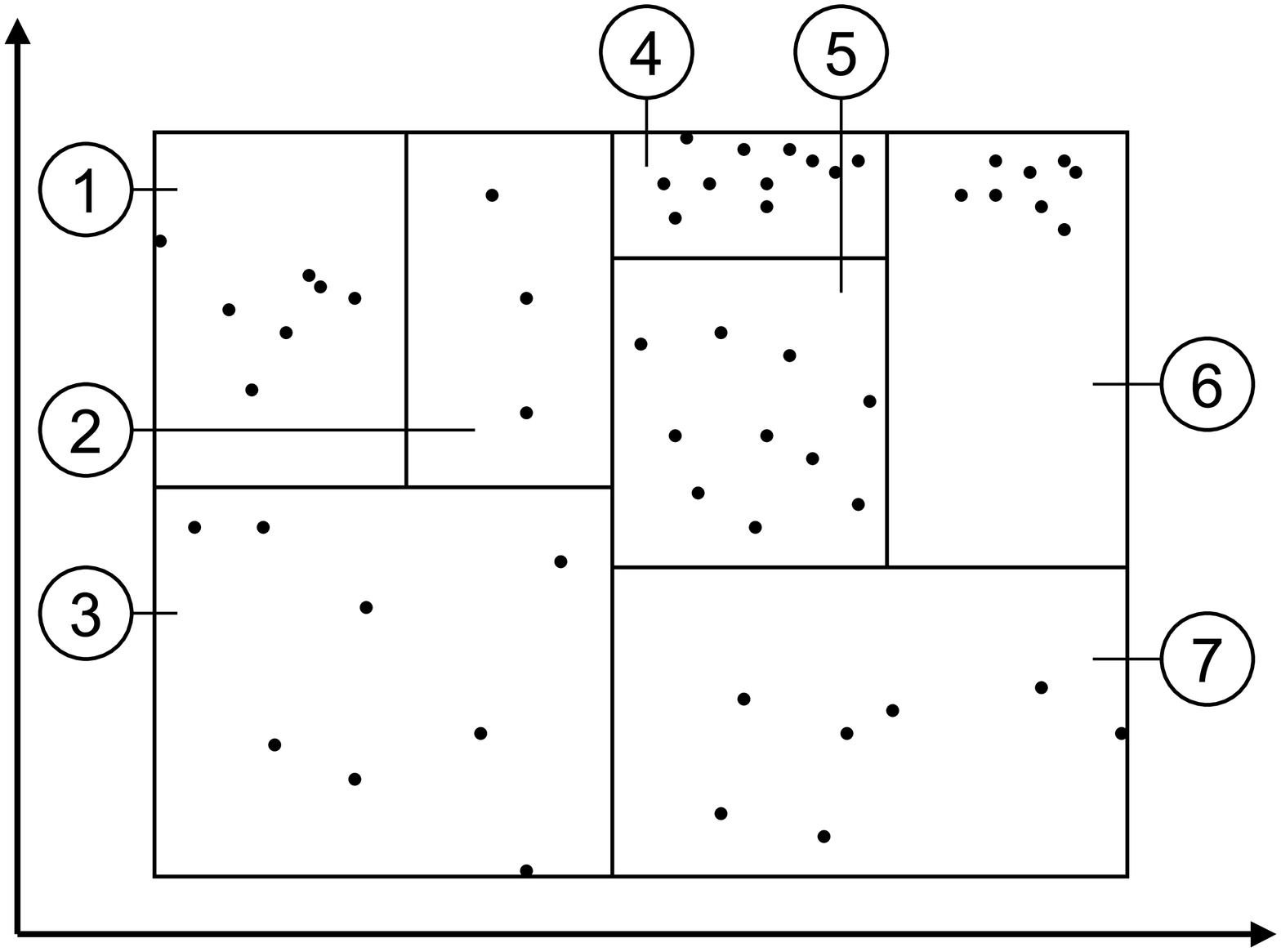} &
\includegraphics[width=1.25in,angle=90]{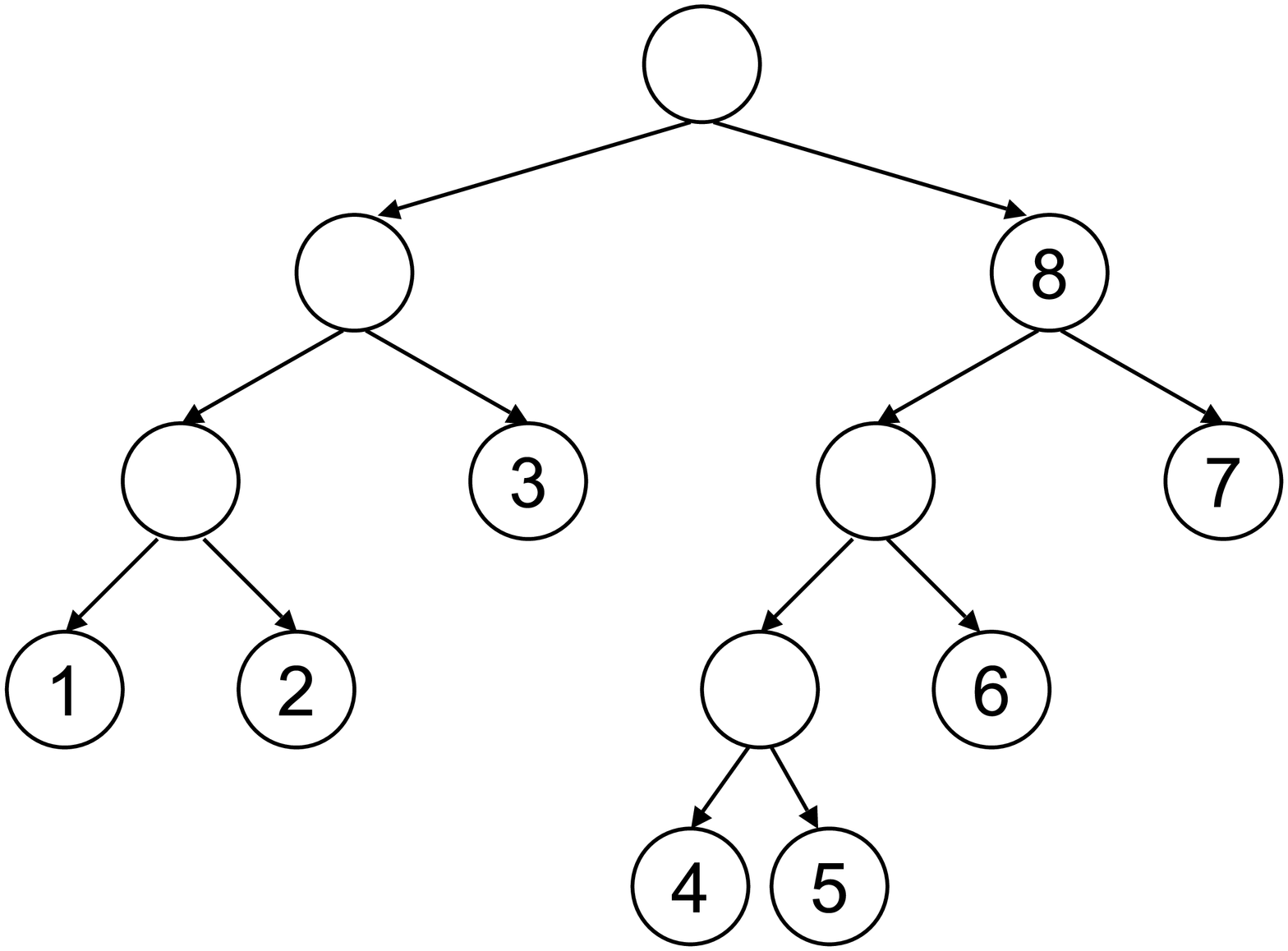} &
\includegraphics[width=1.25in,angle=90]{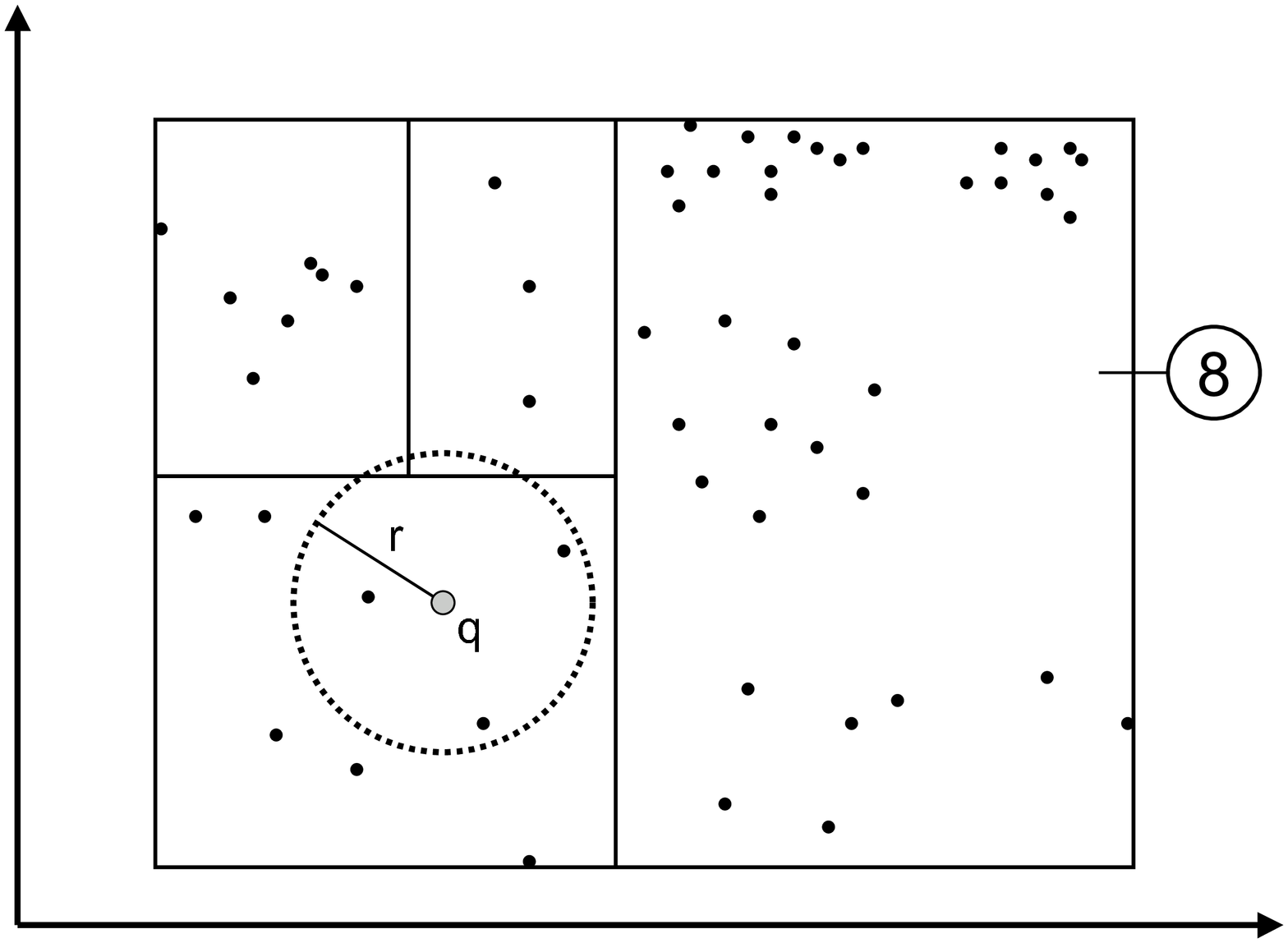} \\
(A) & (B) & (C)
\end{tabular}
\end{center}
\caption{\label{fig:kdtree} A kd-tree built from a set of two
dimensional points (A and B).  During a spatial search we can use the
tree's structure to prune entire subsets of points that \emph{cannot}
fall within proximity $r$ of the query point ${\mathbf q}$, such as
\emph{all} of the points owned by node 8 (C).}
\end{figure}

The hierarchical structure of the tree-based data structures can make
spatial queries very efficient.  Consider the \emph{range search}
query shown in Figure~\ref{fig:kdrangesearch}, where the goal is to find all
points that fall within some radius $r$ of a given query point
${\mathbf q}$.  We simply descend the tree in a depth first search and
look for data points within $r$ of ${\mathbf q}$.  If we reach a leaf
node, we explicitly test the points owned by that node to determine if
their distance from ${\mathbf q}$ is less than $r$.  If so, we add
them to our list of results.  However, we can exploit the spatial
structure to stop exploring a branch of the tree if we find that no
point contained in that branch could fall within our search radius.
For example, in Figure~\ref{fig:kdtree}C we can prune the sub-tree at
node 8 because the entire node falls outside of our search radius.
Thus, we do not have to explore any of node 8's children or test their
associated points.  The ability to prune unfeasible
regions of the search space provides significant computational
savings.

\subsection{Intra-night linking}
\label{ss.IntraNightLinking}

We can extend the spatial query described above to look for simple
intra-night associations by incorporating the temporal aspect of the
data into the search.  Specifically, we do this using a form of
sequential track initiation.  \citep[For a good introduction
see][]{Bar95,Bar01,Bla99}.  We start with an initial trajectory estimate
for the tracklet at some time step and sequentially consider the
subsequent time steps, looking for later detections to confirm,
extend, and refine the tracklet.  In the case of intra-night linkages,
we are starting from individual point detections and thus an
incomplete estimate of the tracklet.

Formally, we consider each individual detection as the start of a
potential tracklet and look for detections at subsequent time steps to
confirm and estimate the tracklet.  We can limit the valid initial
pairings by placing a reasonable restriction on velocities based on
our estimate of {\it a priori} velocity distributions or trailing
information.  For each valid match we use the pair of detections to
define the tracklet and then search later time steps for other
consistent detections.  This allows us to confirm the tracklet and
effectively find all detections that belong to a given tracklet.  The
sequential intra-night linkage algorithm is given in
Figure~\ref{fig:sequentialalg}.

\begin{figure}
\begin{center}
\begin{tabular}{cc}
\includegraphics[width=2.0in,angle=90]{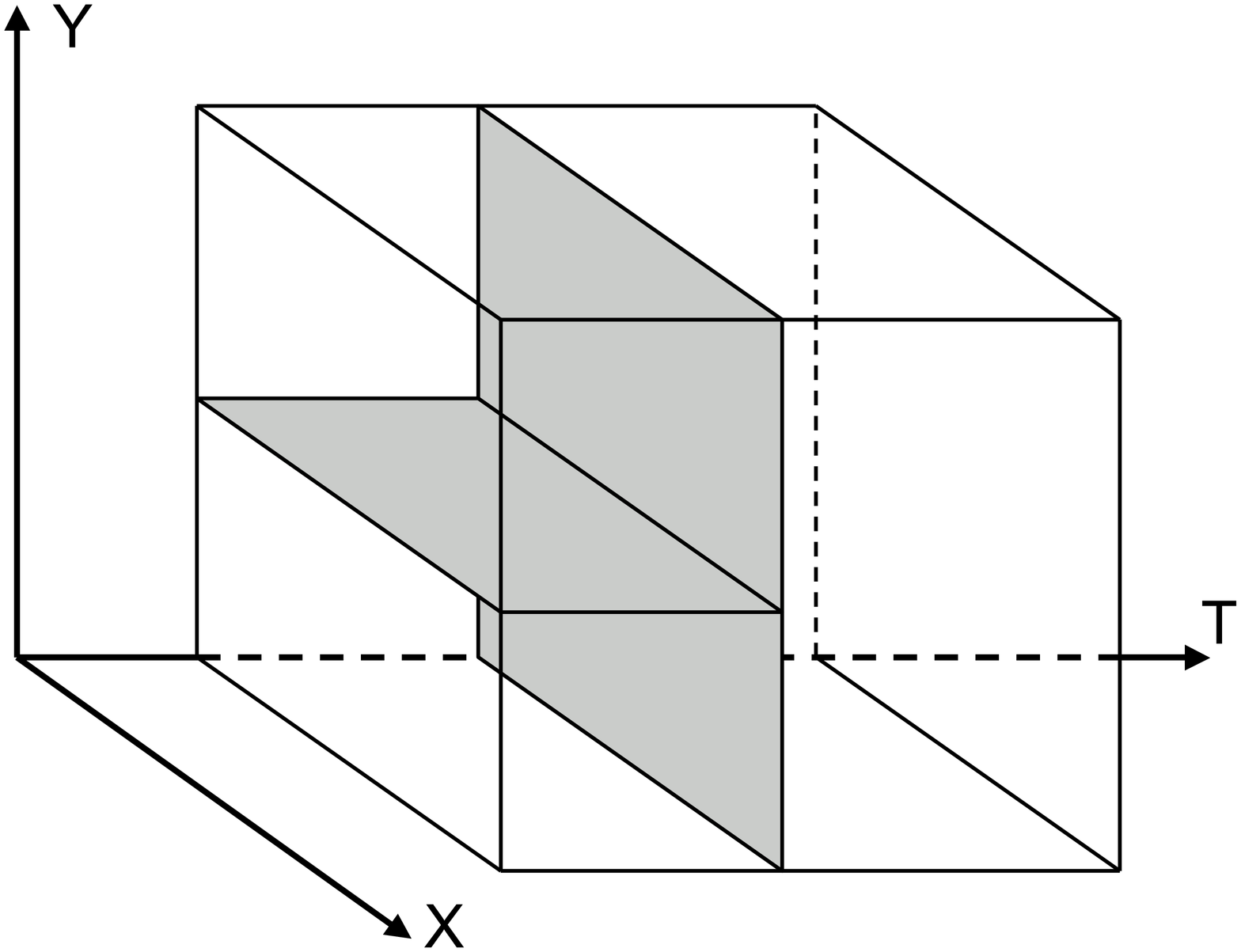} &
\includegraphics[width=2.0in,angle=90]{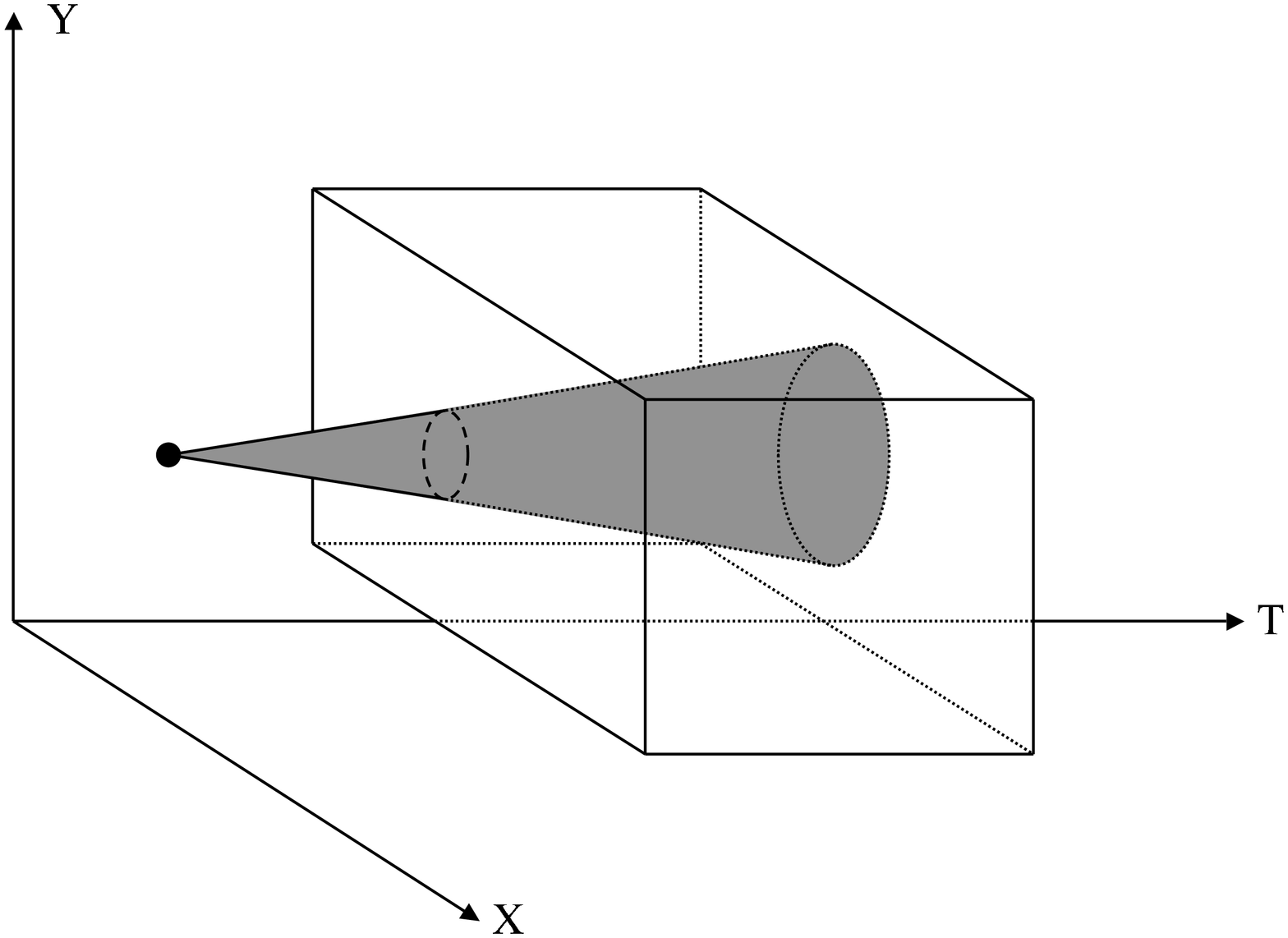} \\
(A) & (B)
\end{tabular}
\end{center}
\caption{ \label{fig:movingobjrange} We can add time as a dimension to
the kd-tree, partitioning the data in both space and time (A).  Given
a kd-tree constructed on both position and time, the moving object
range search is equivalent to searching a cone out from the query
point where the spread of the cone is controlled by the maximum
allowed speed (B).}
\end{figure}

In order to perform the linking efficiently in large scale domains, we
employ the kd-tree with both spatial and temporal structure in the
search.  As shown in Figure~\ref{fig:movingobjrange}A, we can do this
by constructing a single 3-dimensional kd-tree on all of the points by
including time as a dimension.  Given this tree we can then
efficiently search for both the first pairing and the later confirming
detections, by extracting only those detections that are reachable
given our query point and velocity bounds.  As shown in
Figure~\ref{fig:movingobjrange}B, this query effectively searches a
cone projecting out from the query point ${\mathbf q}$.  The algorithm
for finding the feasible points, shown in
Figure~\ref{fig:movingobjsearchalg}, is a range search centered on
${\mathbf q}$'s position.  Unlike the standard kd-tree range search,
we define the range with respect to the current node's time bounds
[$t_{min}$, $t_{max}$] and the overall velocity bounds [$v_{min}$,
$v_{max}$].  We can prune the search if no point in the current node
is reachable from ${\mathbf q}$ given the velocity bounds.

Given a query point ${\mathbf q}$ at time $t_q$ such that
$t_q < t_{min}$, we can prune if:
\begin{equation} \label{eq:movingrangeprune1}
MIN \,\, dist({\mathbf q}, {\mathbf y} )_{{\mathbf y} \in node} > v_{max} \cdot (t_{max} - t_q)
\end{equation}
or
\begin{equation} \label{eq:movingrangeprune2}
MAX \,\, dist({\mathbf q}, {\mathbf y} )_{{\mathbf y} \in node} < v_{min} \cdot (t_{min} - t_q)
\end{equation}
where $dist({\mathbf q}, {\mathbf y})$ represents the distance between
the points $\mathbf q$ and $\mathbf y$.  An analogous pruning rule
applies for cases where $t_q > t_{max}$.  In the above tests,
${\mathbf y}$ does not have to be an actual data point.  Rather
${\mathbf y}$ can be any point within the node's bounding box.

We also incorporate trailing information, if available, into the
algorithm both to limit the search for associations and to filter the
proposed tracklets.  First, we use information about the length of
the detection and the exposure time to estimate the object's angular
velocity.  This estimate, along with its associated error, is used
to define the object's minimum and maximum possible velocity, allowing
us to adapt the search to each individual detection.  Second, we use
the trail's orientation (and its associated error) to filter the
proposed tracklets by requiring that all detections in the tracklet
have similar orientations.  When the length of a trail is sufficiently
small the trail's length and angle become unreliable and the trail is
ignored; i.e. the trail is treated as a point-source detection.

The intra-night linking algorithm described here did not use
photometric information when creating tracklets.  This is due to the
fact that the vast majority of all detections will be close to the
system's limiting magnitude and therefore in a photometric regime
where large statistical errors are present.  The constraints offered
by checking the photometry are weak and, we will show below, our
simulations suggest that it is unnecessary - we obtain high efficiency
and sufficient accuracy to allow the system to operate well without
taking photometry into account.  It will be trivial to implement a
constraint on consistent photometry between detections if we find it
necessary to do so after further study.

\subsection{Inter-night linking}
\label{ss.InterNightLinking}

The primary benefit of spatial data structures is the ability to prune
and thus ignore regions that are ``obviously'' infeasible given
our query.  We can extend this notion to finding associations, and
thus new tracklets or tracks, by explicitly searching for entire sets
of points that are mutually compatible \citep{Kub05a,Kub05b}.
The primary benefit of searching for entire sets of points is that we
can often avoid many early dead-ends that may result from trying to
establish the first few associations in a track.  Specifically, many
pairs of tracklets may look like promising matches, but be left
unconfirmed by later supporting detections.  In fact, the problem of
many good initial pairings becomes significantly worse as the gap in
time between observations of the same object increases.

\subsubsection{Searching Sets of Model Points}

This process can be summarized as: given two or more regions (bounding
both position and possibly velocity) at different times is there a
track that can pass through them?  If so, are there other points that
would confirm this track?

We can identify potential tracks by searching over all sets of
tracklets that \emph{could} define the track.  In the case of
inter-night linking with quadratic tracks (in motion in both Right
Ascension and declination) we can search over all pairs of tracklets
that could be used to define a quadratic and then check for additional
supporting tracklets to confirm these proposed tracks.  The benefit of
such an approach is that we can quickly search the models defined by
the data and efficiently test whether these models are supported.
Again, we can do this search efficiently by using spatial data
structures such as kd-trees.

In order to efficiently search over all sets of points or tracklets
that could define a valid model, we want to be able to use spatial
structure from all the points, including those at different time
steps.  We can do this by building \emph{multiple} kd-trees over
detections (one for each time step) and searching combinations of tree
nodes.  At each level of the search, our current search state consists
of a \emph{set} of tree nodes that define areas in which the track
could be at those time steps.  Thus we are effectively saying: ``One
of the points in the set could be owned by the first tree node,
another could be owned by the second tree node, etc.''  As the search
descends, each of the nodes' bounding boxes shrink, limiting the areas
in which the track could occur and thus zeroing in on track positions
at each time.  At the limit, the search reaches a set of individual
detections (from different time steps) that are all mutually
compatible with a single track.  We can also use the same approach for
linking tracklets by treating the tracklet's velocity as two
additional dimensions.

\begin{figure}[!tb]
\begin{center}
\includegraphics[width=0.3\textwidth,angle=90]{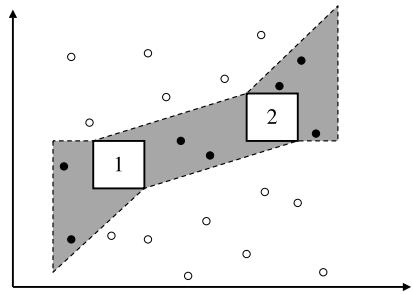} 
\end{center}
\caption{The model nodes' bounds (1 and 2) define a region of feasible
support (shaded) for \emph{any} combination of model points from those
nodes.}
\label{fig:pointlist}
\end{figure}

For example, in the simple linear case the model is defined by only 2
points, thus we can efficiently search through all possible models
using 2 \emph{model} nodes to represent the current search state.  At
each stage in the search we are effectively considering all possible
models that could be formed with a point in each of our two tree
nodes.  In addition, as shown in Figure~\ref{fig:pointlist}, the
spatial bounds of our current model nodes immediately limit the set of
feasible support points for \emph{all} line segments compatible with
these nodes.  Thus it may be possible to track which support points
are feasible and use this information to prune the search due to a
lack of support for \emph{any} model defined by the points in those
nodes.

\subsubsection{Variable-trees algorithm}

The variable-tree algorithm works by searching over all sets of points
that could define a model while tracking which points could support
the current set of models.  As described above, the algorithm uses a
multiple tree search over model defining points to close in on valid
models.  In addition, throughout the search we track which points
could support our current set of models using an adaptive, dynamic
representation of the points in the support space.

\begin{figure}[!tb]
\begin{center}
\includegraphics[width=0.3\textwidth,angle=90]{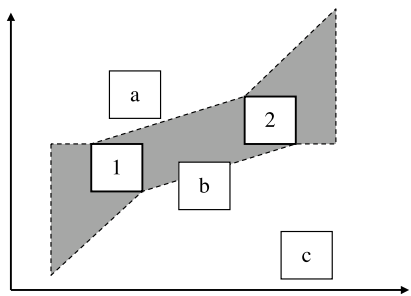}
\end{center}
\caption{The model nodes' bounds (1 and 2) define a region of feasible
support (shaded) against which we can classify entire support tree
nodes as feasible (node b) or infeasible (nodes a and c).}
\label{fig:nodelist}
\end{figure}

The key idea behind the variable-tree search is that we can use a
\emph{dynamic} representation of the potential support.  Specifically,
we can place the support points in trees and maintain a dynamic
\emph{list} of currently valid support nodes.  As shown in
Figure~\ref{fig:nodelist}, by only testing entire nodes (instead of
individual points), we are using spatial coherence of the support
points to remove the expense of testing each support point at each
step in the search.  And by maintaining a list of support tree nodes,
we are no longer branching the search over these trees.  Thus we
remove the need to make a hard ``left or right'' decision.  Further,
using a combination of a list and a tree for our representation allows
us to refine our support representation on the fly.  If we reach a
point in the search where a support node is no longer valid, we can
simply drop it off the list.  And if we reach a point where a support
node provides too coarse a representation of the current support
space, we can simply remove it and add both of its children to the
list.

The primary advantage of this search approach is that it allows us to
use structure from all aspects of the problem.  We are able to test
entire sets of supporting points against entire sets of models,
removing the need to test a huge number of individual combinations.
However, we still maintain the ability to use the information provided
by the support points, pruning the search if a model is not supported
by a sufficient number of additional detections.  Further, by
adaptively changing our representation, we can balance the testing
cost and the pruning power of the search.

\begin{figure} [!tb]
\begin{center}
\begin{tabular}{lll}
Search Step 1:& & \\
\includegraphics[width=1.25in,angle=90]{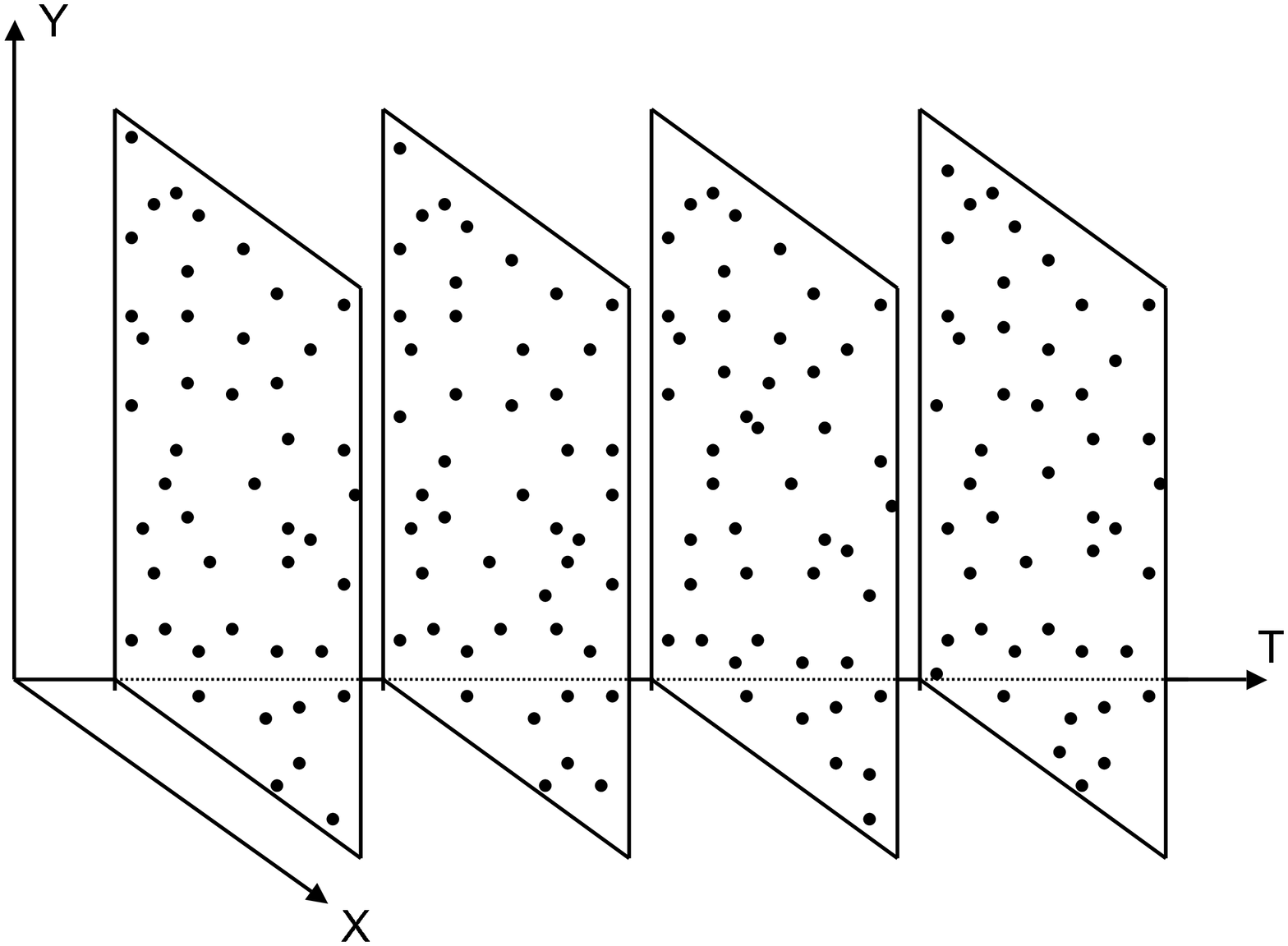} &
\includegraphics[width=1.25in,angle=90]{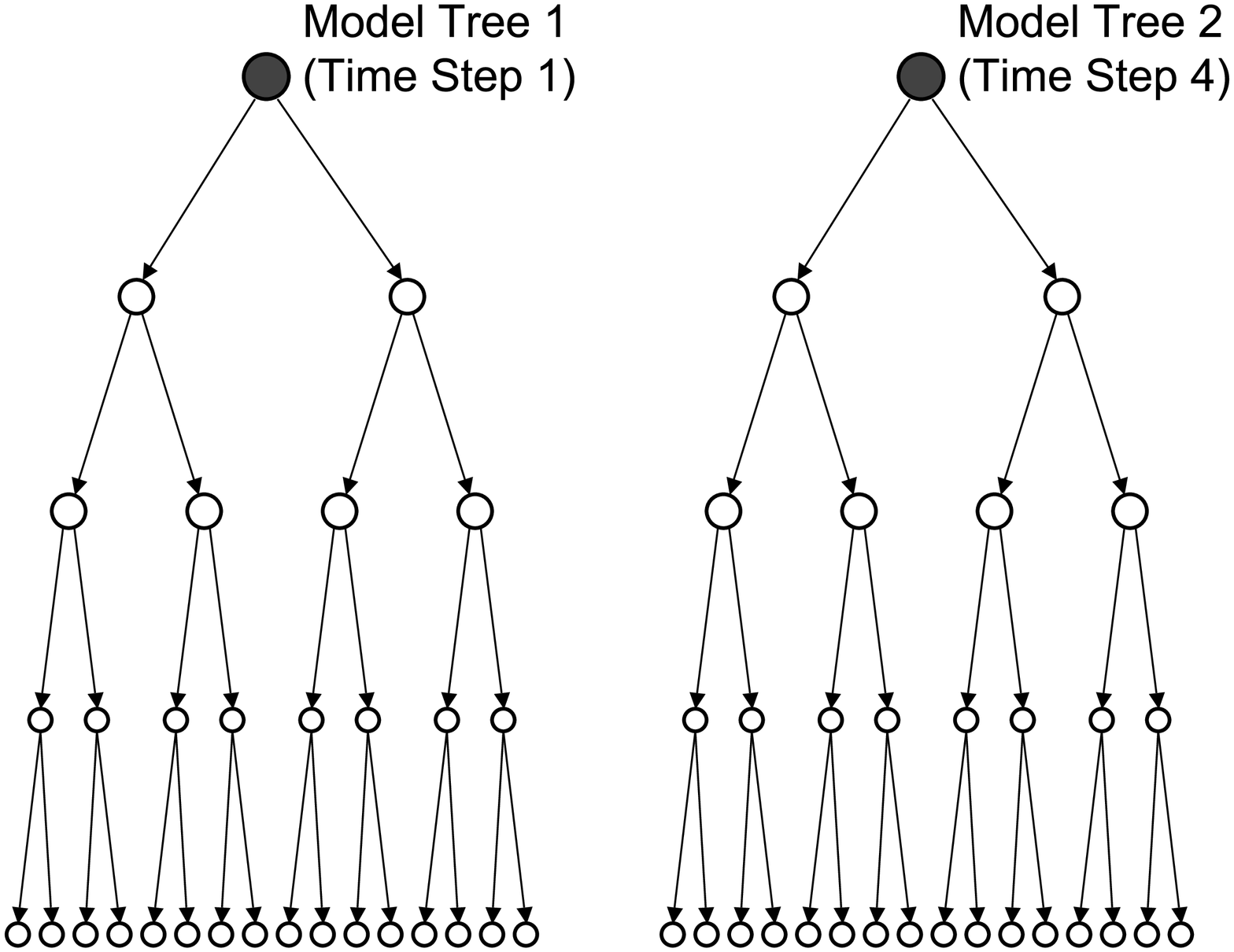} &
\includegraphics[width=1.25in,angle=90]{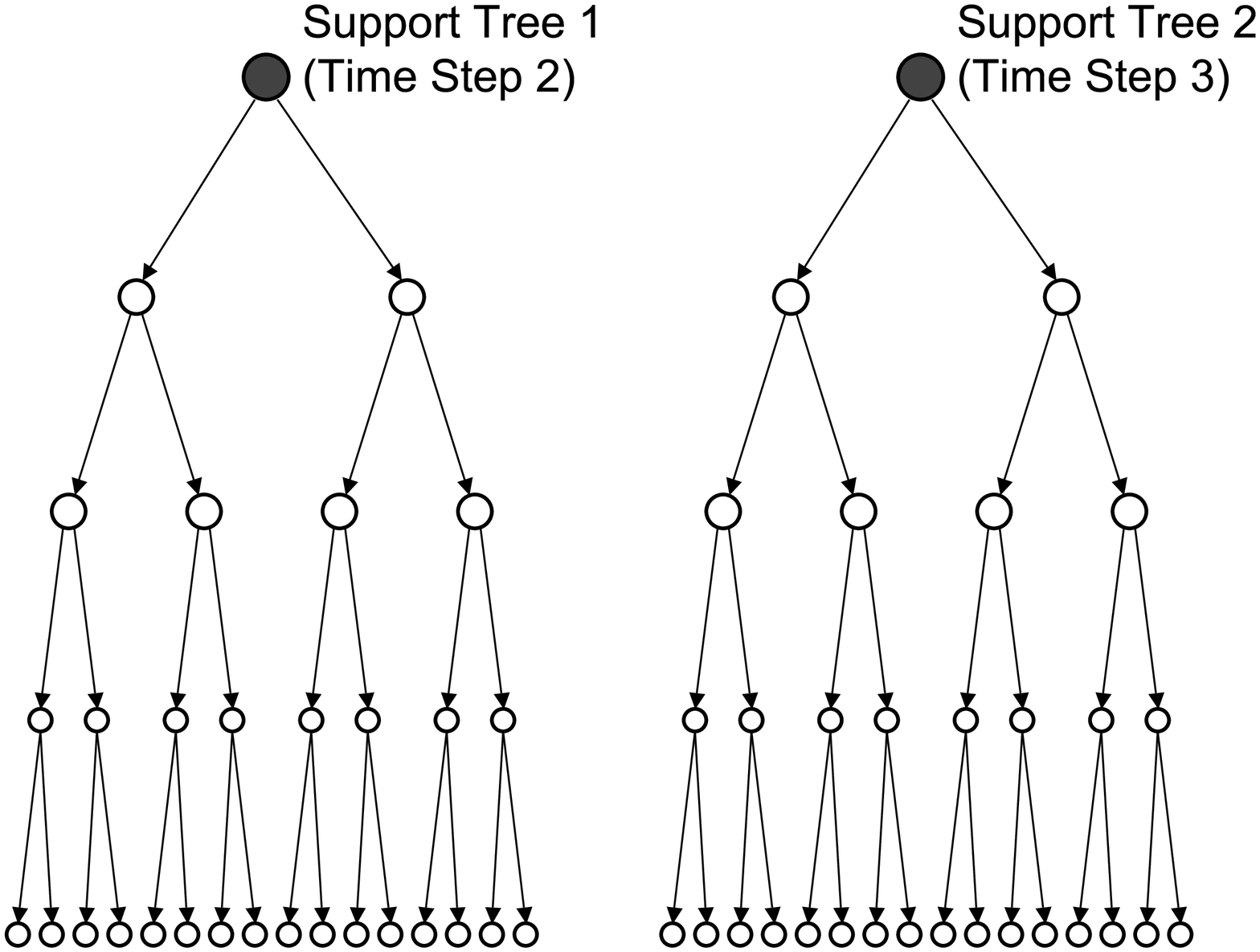} \\
\hline
Search Step 2:& & \\
\includegraphics[width=1.25in,angle=90]{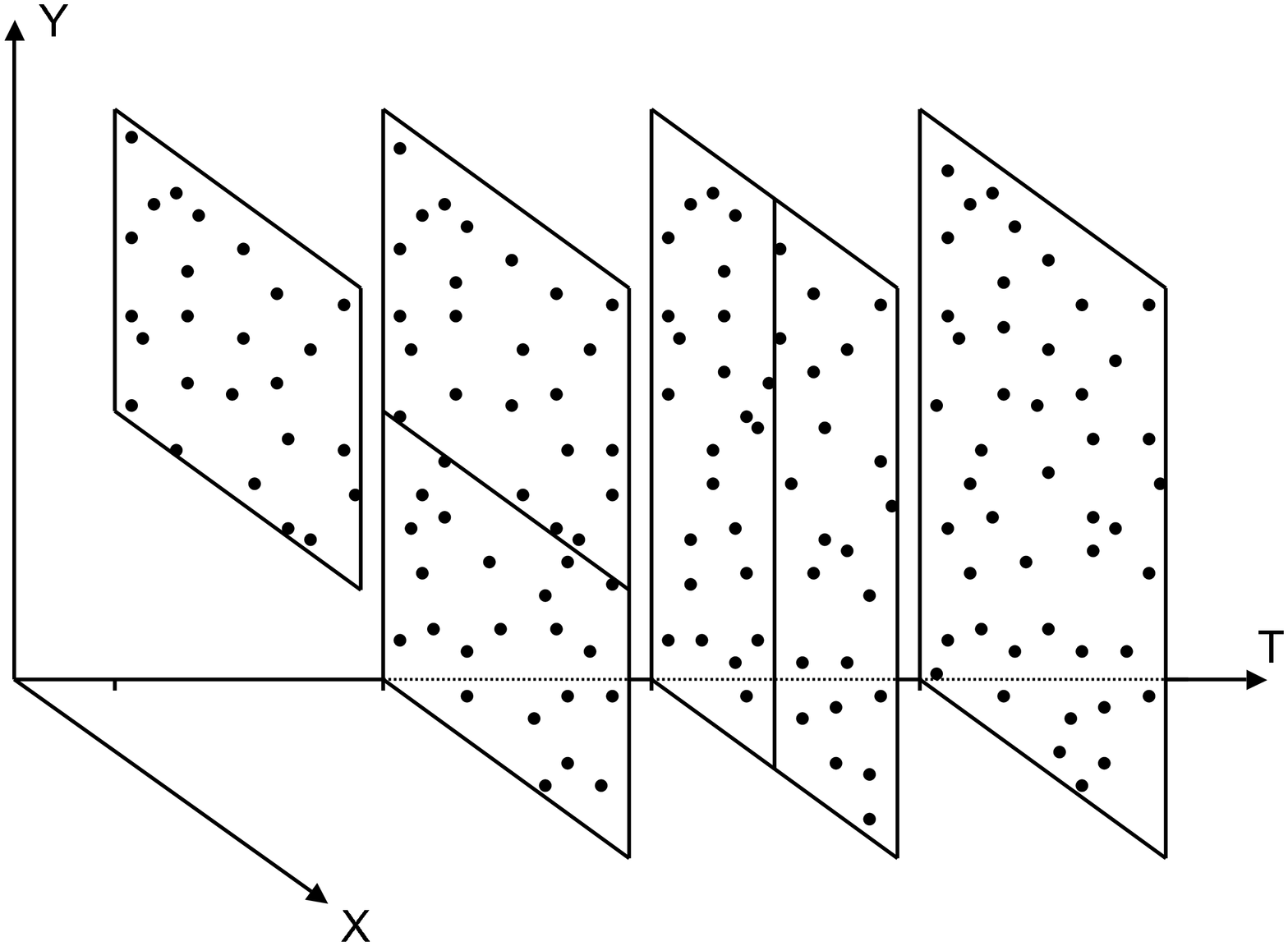} &
\includegraphics[width=1.25in,angle=90]{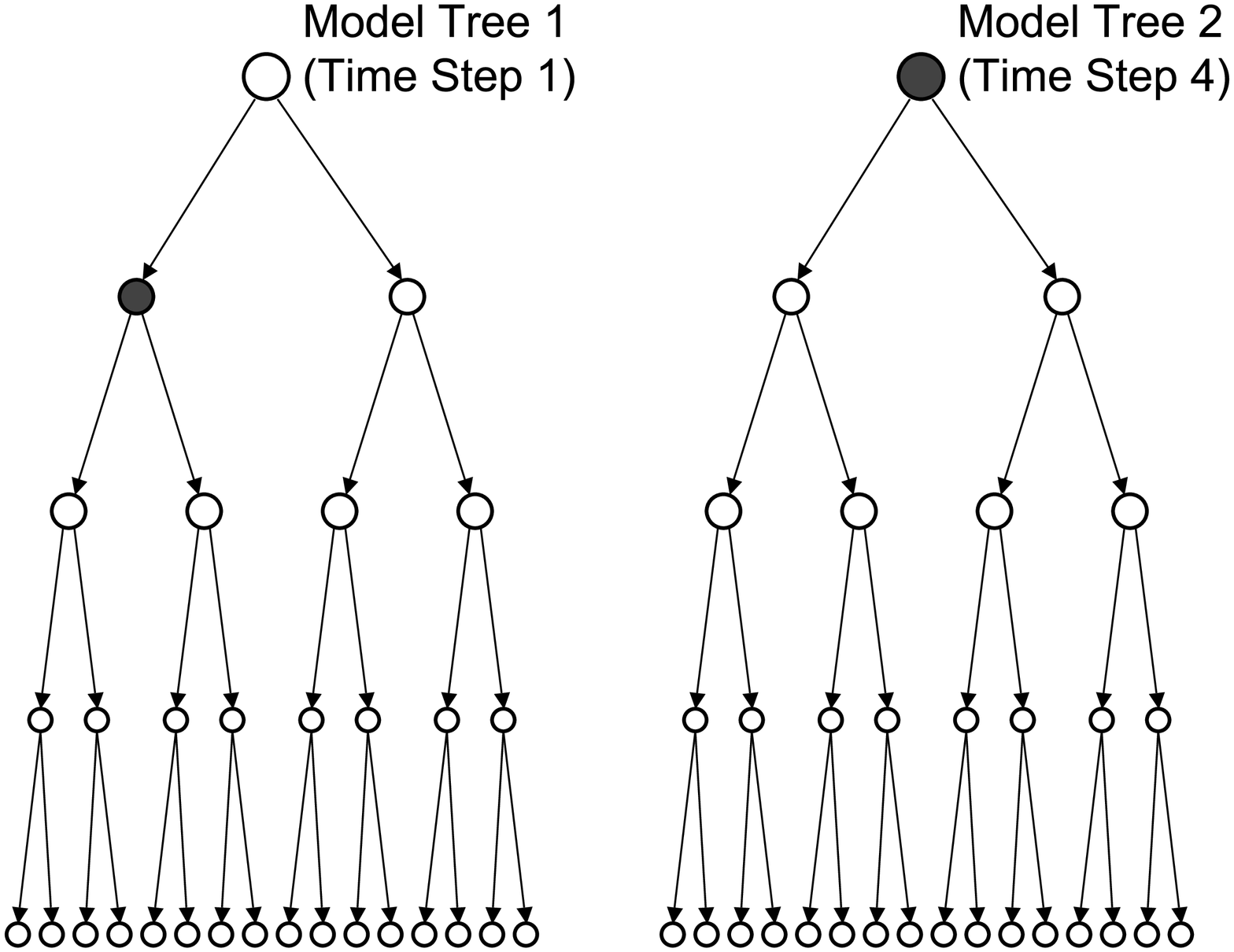} &
\includegraphics[width=1.25in,angle=90]{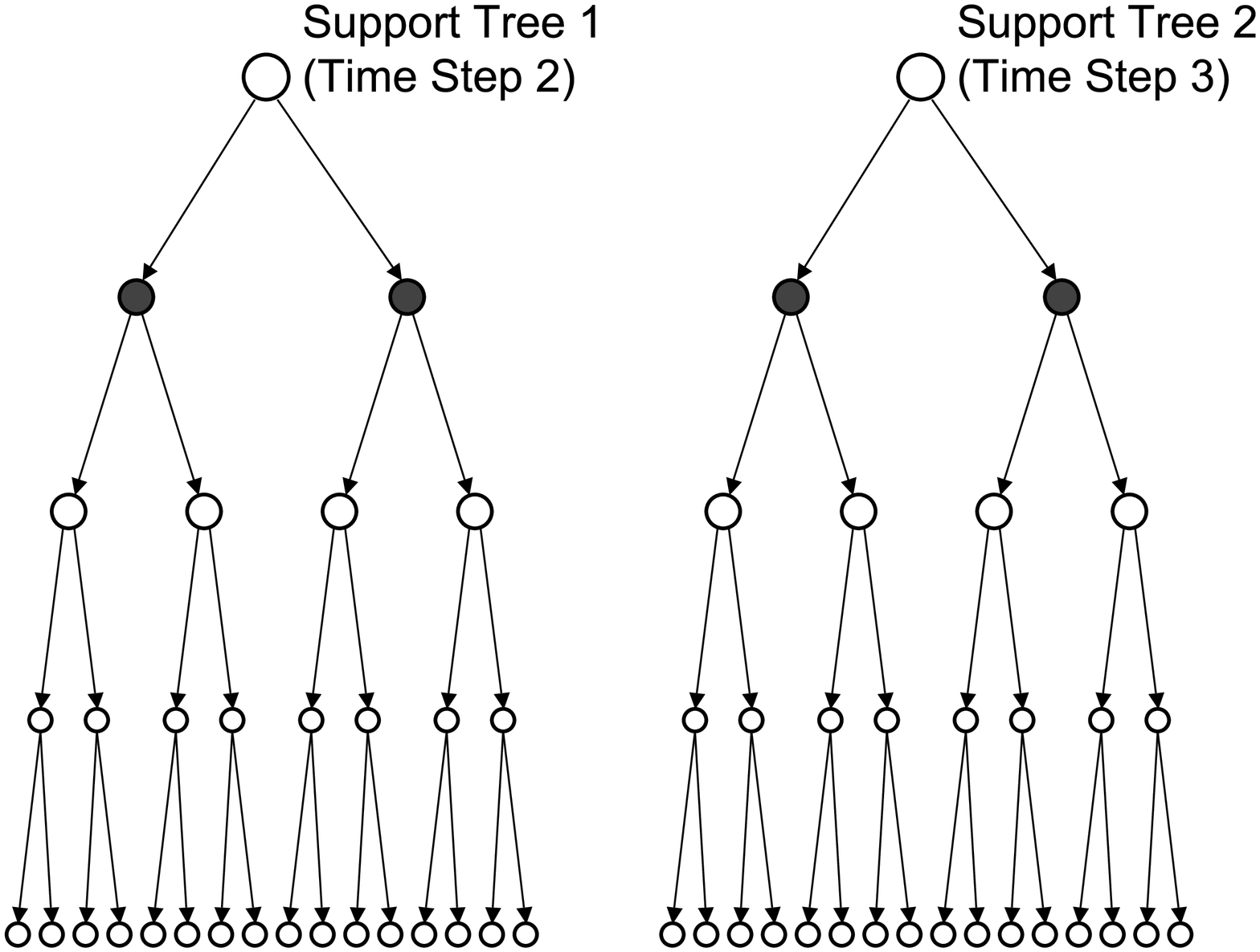} \\
\hline
Search Step 3:& & \\
\includegraphics[width=1.25in,angle=90]{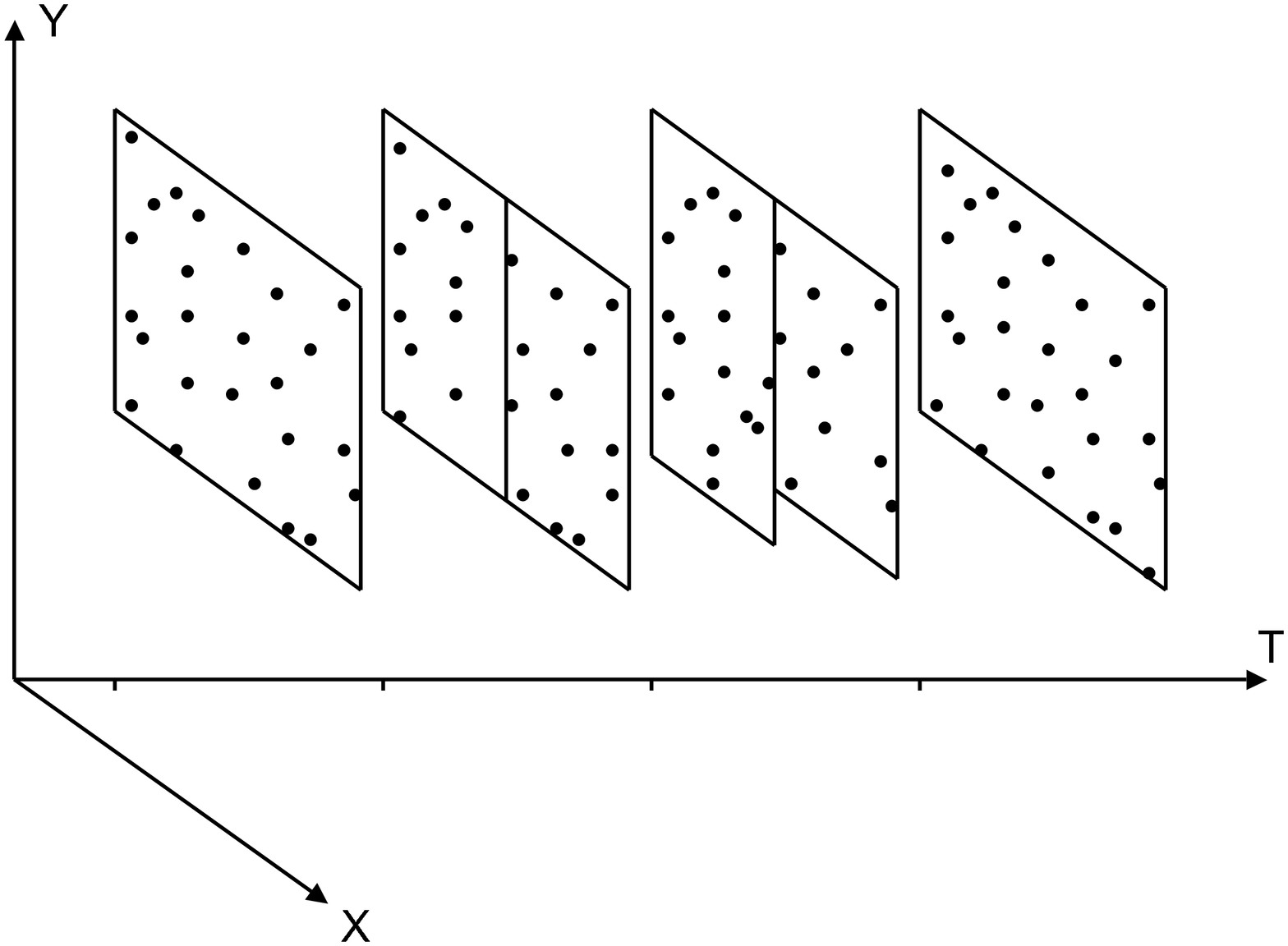} &
\includegraphics[width=1.25in,angle=90]{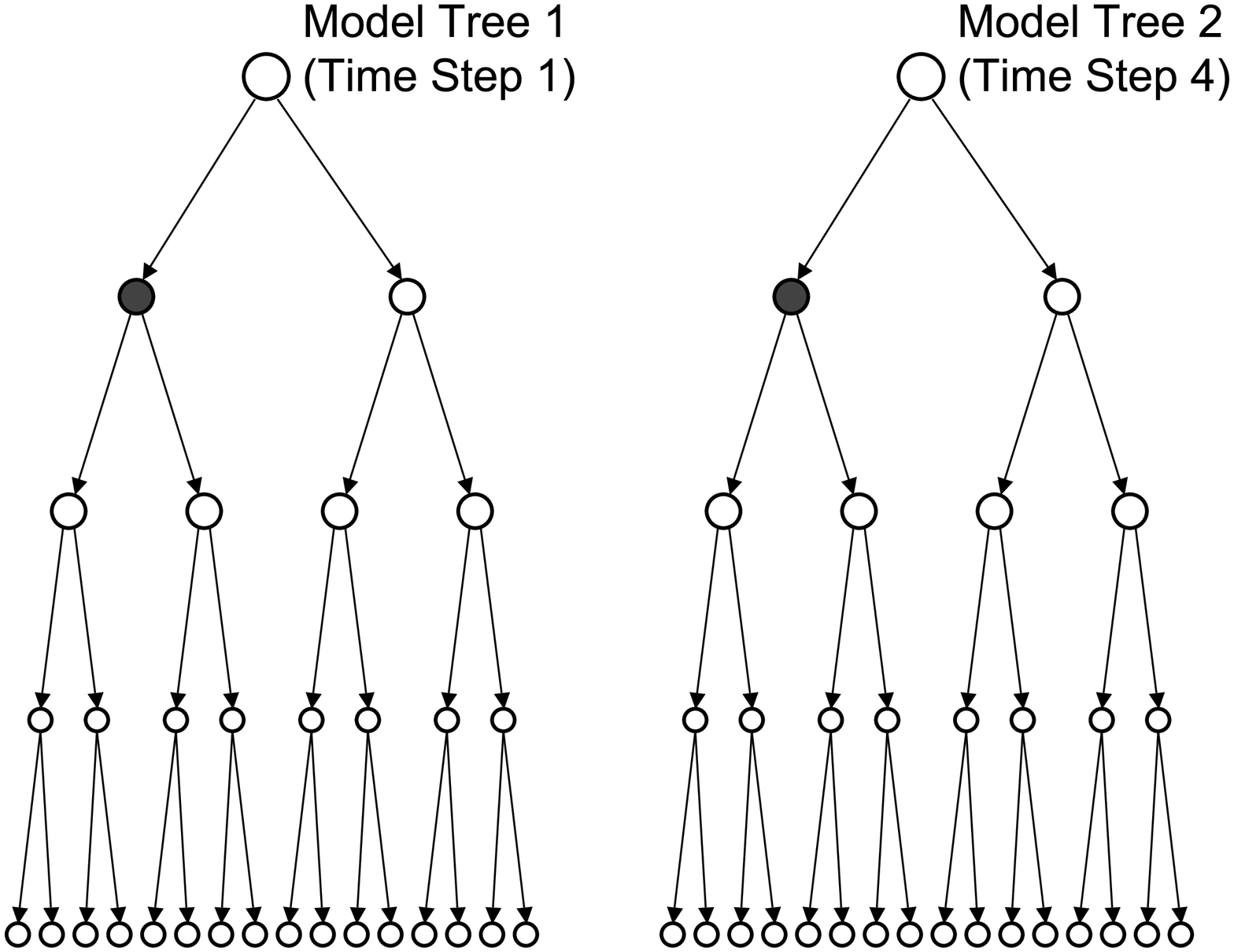} &
\includegraphics[width=1.25in,angle=90]{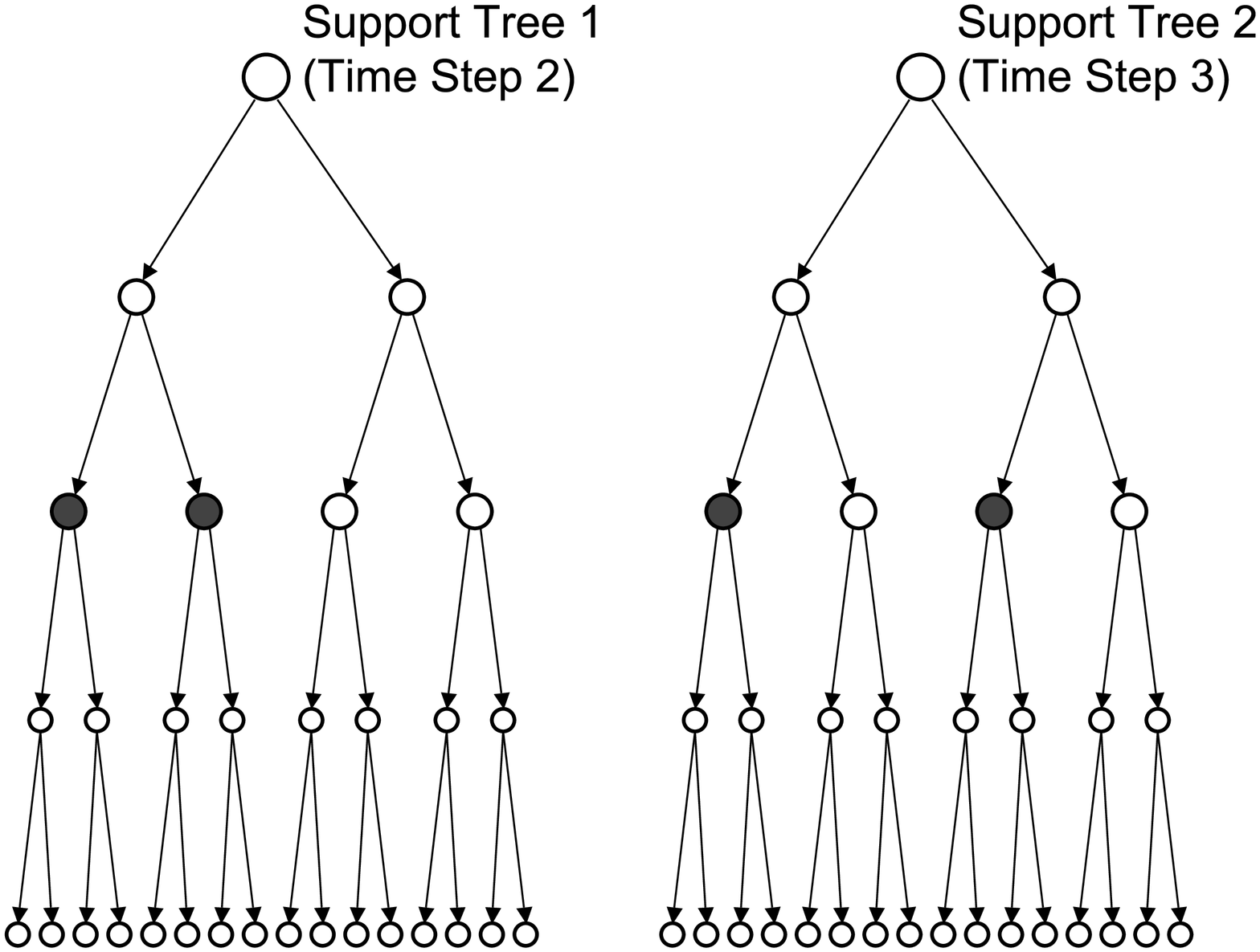} \\
\hline
Search Step 5:& & \\
\includegraphics[width=1.25in,angle=90]{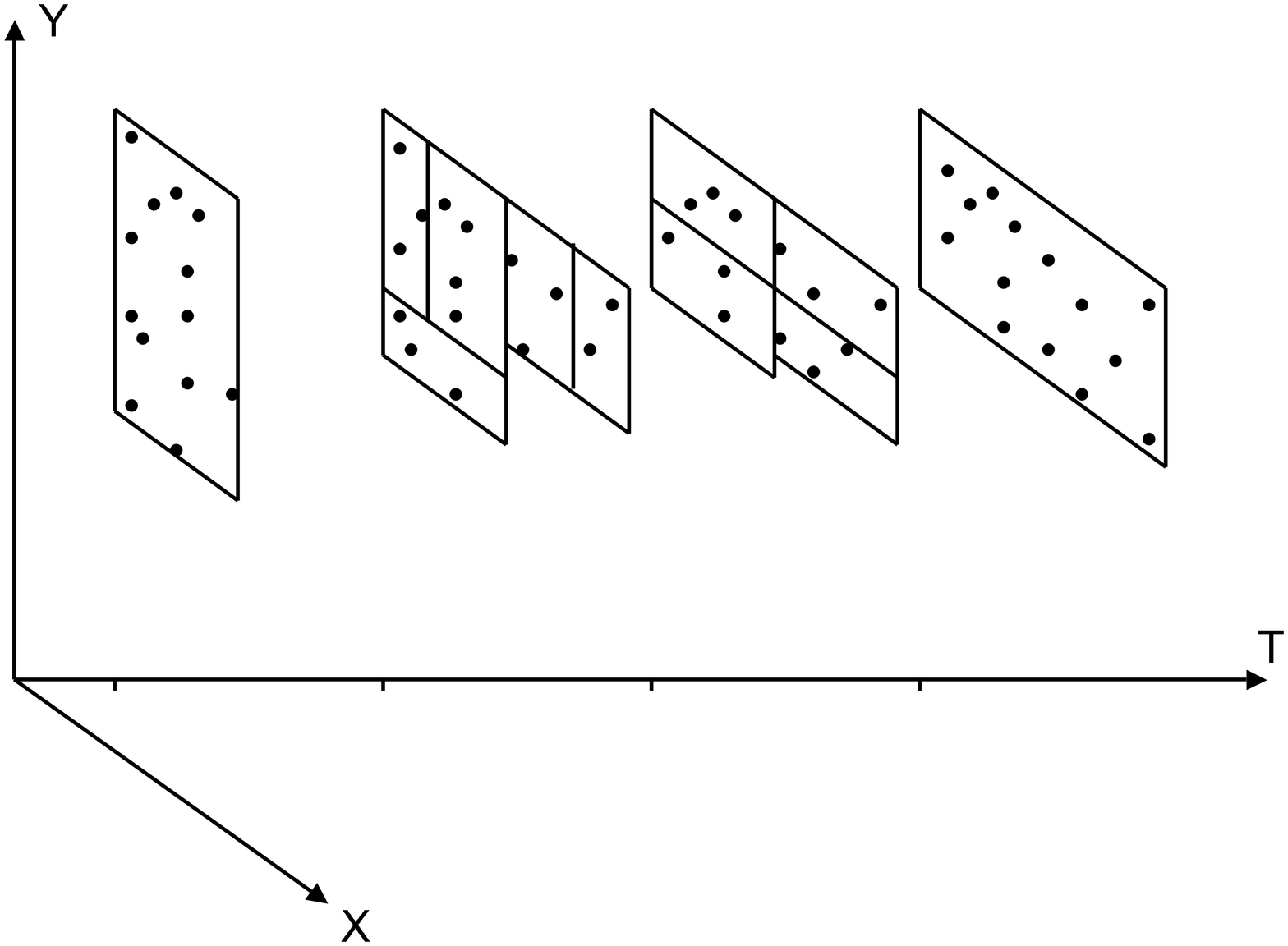} &
\includegraphics[width=1.25in,angle=90]{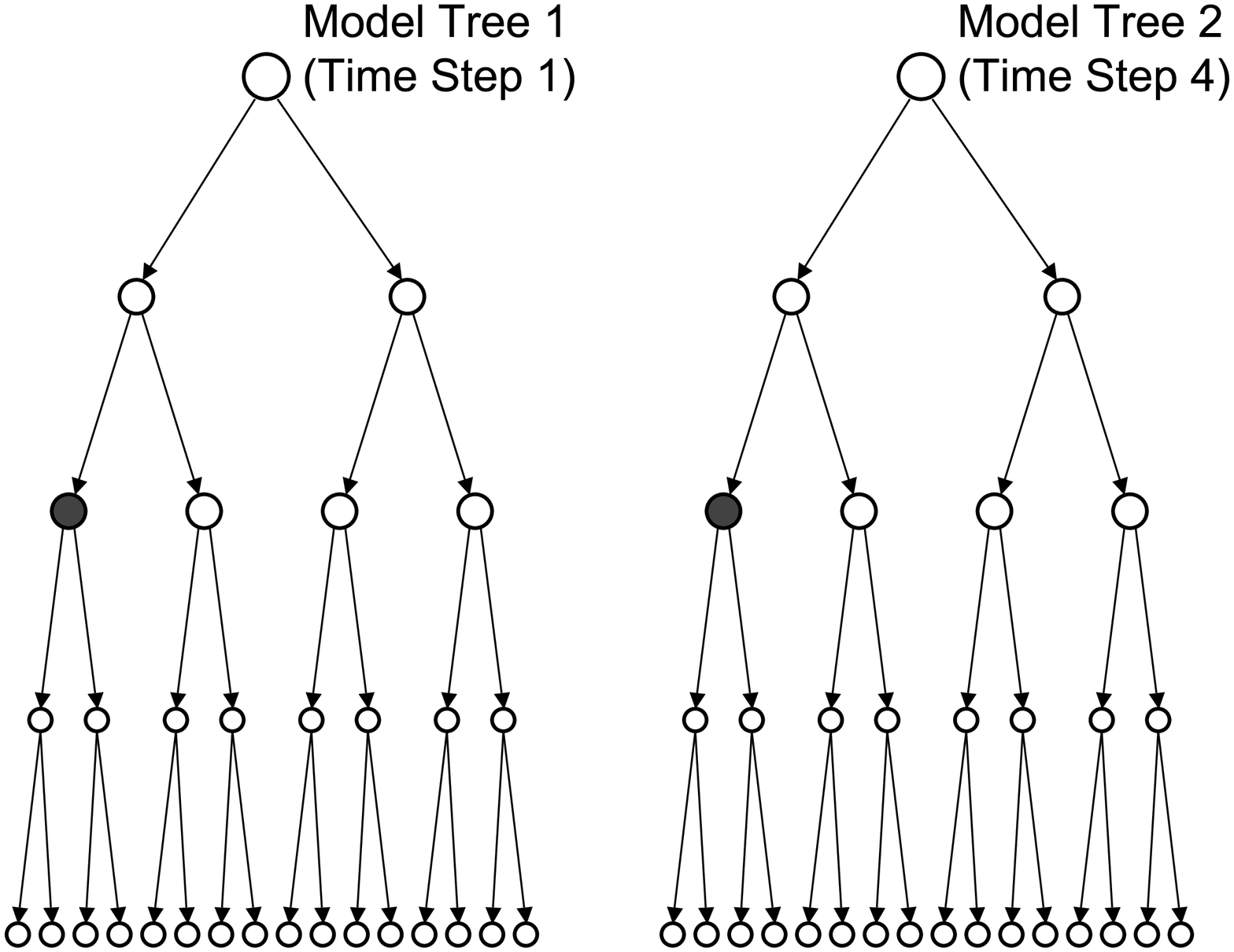} &
\includegraphics[width=1.25in,angle=90]{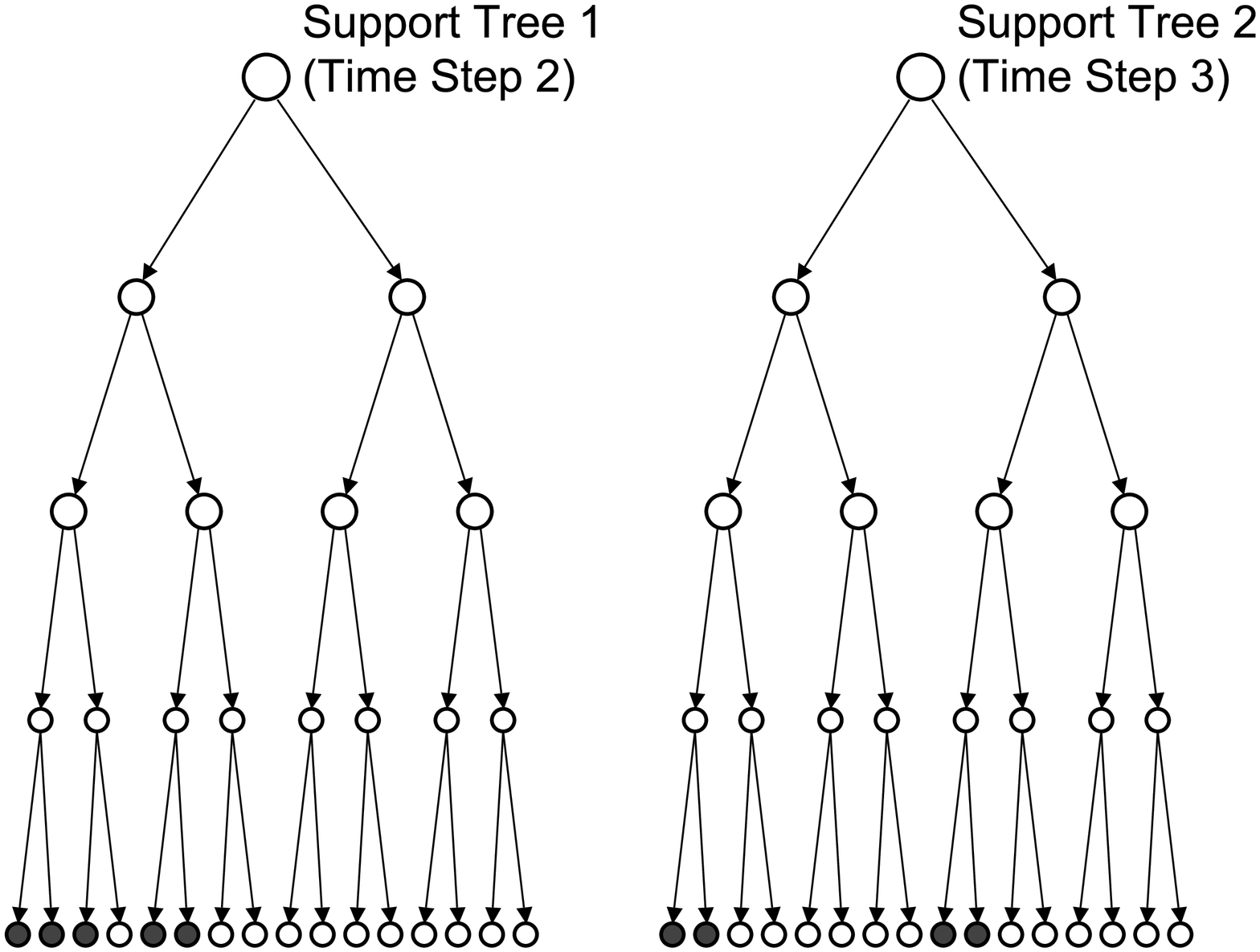} 
\end{tabular}
\end{center}
\caption{The variable-tree algorithm looks for valid tracks by
performing a depth first search over the model trees' nodes.  At each
level of the search the model tree nodes are checked for compatibility
with each other and the search is pruned if they are not compatible.
In addition, the algorithm maintains a list of compatible support tree
nodes.  Since we are not guiding the search with the support trees we
can split the support trees and add: the right child, the left child,
both children, \emph{or} neither child to our list of support tree
nodes.  This figure shows a simple rule where the support tree nodes
are split exactly once at each level of the search.  Support tree
nodes are only added if they are compatible with the entire set of
model tree nodes.  The intermediate step that would be Step 4 has been
intentionally left out.} \label{fig:vtreesearch}
\end{figure}

The full variable-tree algorithm is given in
Figure~\ref{fig:multtreealg}.  A simple example of finding
\emph{linear} tracks while using the track's endpoints (earliest and
latest in time) as model points and using all other points for support
is illustrated in Figure~\ref{fig:vtreesearch}.  The first column
shows all the tree nodes that are currently part of the search.  The
second and third columns show the search's position on the two model
trees and the current set of valid support nodes respectively.  Again,
it is important to note that by testing the support points as we
search, we are both incorporating support information into the pruning
decisions and ``pruning'' the support points for entire sets of models
at once.

In the case of linking tracklets we are also interested in using
bounds on the tracklet's velocity.  The algorithm does this by
treating the tracklets as 5-dimensional points with two angular
positions, two angular velocities, and a time.  These dimensions are
used in constructing and pruning the kd-trees but otherwise do not
affect the algorithm.

\section{Results \& Discussion}
\label{s.ResultsDiscussion}

Our MOPS implementation strategy has been to quickly develop a
prototype system framework for testing purposes that roughly
implements all features of a fully functional system.  Once the
prototype was developed we could examine the efficiency of each MOPS
subsystem and identify bottlenecks in the processing of moving object
detections.  The algorithms described in \S\ref{ss.IntraNightLinking}
and \S\ref{ss.InterNightLinking} for tracklet and track creation have
been implemented and tested on many synthetic models and some real
asteroid survey data.

\subsection{Tracklet Identification}
\label{ss.TrackletIdentification}

The tracklet identification algorithm is known as
\verb#findTracklets#.  It is called after all fields have been
acquired on a night and operates on all detections from the difference
images (\ie after static-sky subtraction).  It might be argued that
\verb#findTracklets# should be invoked for each pair of images
separated by a Transient Time Interval (TTI) but this would create
separate tracklets for detections of objects in the overlapping areas
of adjacent fields.

As described above, \verb#findTracklets# accumulates detections from a
single night into tracklets consistent with linear motion through the
night.  It is a `greedy' algorithm in that it always tries to group
the maximum number of detections into a tracklet consistent with the
limits on motion and astrometric position.  In practice, we have found
that we need to limit the algorithm to accumulating detections into
tracklets to those that are separated by less than a critical
threshold time set to a few times the TTI typical of the re-visit time
for a field on a single night (we are using a one hour limit for the
formation of tracklets at the moment).  The upper limit to the time
difference between the detections in a tracklet means that
`intra-night' linking does not necessarily link together all available
observations of an object within a single night.  This situation might
occur, for instance, for an object that appears in a part of the image
that overlaps with an adjacent field that, for one reason or another,
is not acquired within the hour after the first field is imaged.  In
this case our system would generate two separate tracklets for the night.

As the sky-plane density of real and false detections increases we
expect that both the efficiency (percentage of synthetic tracklets
identified) and accuracy (percentage of identified tracklets that are
synthetic) will decrease.  We have found that the performance of the
standard \verb#findTracklets# algorithm is so close to 100\% under all
circumstances and for all types of synthetic solar system objects that
it makes no sense to discuss the results other than gross totals.  The
standard algorithm implements the option described at the end of
\S\ref{ss.IntraNightLinking} of using trailing information for each
detection in order to prune the number of feasible intra-night links.
Table \ref{tab.StandardTrackletPerformance} shows the results we have
achieved in both the opposition and sweet spot regions.  The
efficiency could be increased to 100\% by extending the search radius
for linking detections but this comes at the cost of decreasing the
linking accuracy and increasing the fraction of discordant, mixed and
spurious tracklets (see the Table
\ref{tab.StandardTrackletPerformance} caption for the definition of
these terms).

The final choice of all \verb#findTracklets# parameters will be made
when the entire MOPS is fully functional.  The values will be set in
order to optimize the overall system rather than the efficiency of
accuracy of the intra-night linking.  i.e. It may appear advantageous
to achieve very high operational efficiency for tracklet formation but it
is not clear how the corresponding decreased accuracy and increase in
false tracklets will affect the track formation process (described in
\S\ref{ss.TrackIdentification}).  Of course, under realistic operating
conditions the realized intra-night linking efficiency will be limited
by the fill factor and other operational constraints.

The modest decrease in accuracy and increase in discordant, mixed and spurious
tracklets in the sweet-spots is due to the increase in the mean speed
of objects at small solar elongations.  Since they move faster, the
search radius for intra-night linking needs to be increased and this
has the side effect of increasing the false tracklet rate.

Note that the total number of tracklets available in a lunation is in
excess of 1.3 million corresponding to almost 450,000 different
objects.  Thus, in a single lunation \ps\ may identify and obtain
orbits and colors for more solar system objects than are currently
known.  The PS1 proto-type system with only a single telescope will
not perform as well but will still identify on the order of as many
objects as are currently known in a single lunation.

Table \ref{tab.NonStandardTrackletPerformance} shows the effect of not
using each detection's trailing information when performing
\verb#findTracklets# (see \S\ref{ss.IntraNightLinking} for a brief
discussion of the use of trailing information when creating
tracklets).  As expected, the efficiency can remain high only at the
expense of realizing one-third the accuracy and nearly an order of
magnitude more false tracklets.


\subsection{Track Identification}
\label{ss.TrackIdentification}

The algorithm for inter-night linking of tracklets is called
\verb#linkTracklets#.  Once a night of detections has been processed
by \verb#findTracklets# (\S\ref{ss.TrackletIdentification}) blocks of
(usually contiguous) images in the same region of sky are grouped
together for processing by \verb#linkTracklets# in a `pass' (see table
\ref{tab.TrackIdentificationPerformance}).  A database query
identifies all other tracklets obtained in the surrounding area
(increasing sky-plane distance with time) within the last 14 days and
if there are three available nights for linking within that time frame
then \verb#linkTracklets# attempts to link those tracklets together.

The number of images that may be grouped together depends on the
density of tracklets and the length of time over which inter-night
linking is attempted.  In general, we find that beyond the 14 day
limit the linking algorithm becomes inefficient and inaccurate.
Traversing a large gap in time to look for linkages is prohibitive
because there are too many potential linkages that satisfy the
requirement of quadratic motion and too many real objects are
non-quadratic over the same time period and will not be identified.
We limit the range of acceptable speeds from 0.0$\arcdeg$/day to
10.0$\arcdeg$/day where the lower limit allows us to detect extremely
slow moving distant objects and the upper limit is set by our funding
agency.  The maximum acceleration was set to 0.02$\arcdeg$/day$^2$ in
both RA and declination.

Note that the term `inter-night' linking is strictly not correct due
to the time limit on the spacing of intra-night tracklets as described
in \S\ref{ss.TrackletIdentification}.  Inter-night linking actually
links together all tracklets between {\it and} within nights when
available.

Table \ref{tab.TrackIdentificationPerformance} gives various
performance statistics for the \verb#linkTracklets# algorithm.  To
test the performance as a function of the sky-plane density of objects
we generated four different models as described in the table caption.
The realized sky-plane density of synthetic objects in the field
varies over two orders of magnitude while the tracks that were
available for linking ranged over more than three orders.  In each
case the false detections were kept at the expected density of 250
$deg^{-2}$ for \ps\ operations.

Inter-night linking efficiency decreases slowly with the realized
sky-plane density of synthetic tracks as shown in table
\ref{tab.TrackIdentificationPerformance}.  Even at densities
expected for the full four telescope Pan-STARRS system with a limiting
magnitude of $r \sim 24$mag the track creation efficiency is currently
above 98\%.

Of more concern is the effect of sky-plane density on the accuracy of
track creation - the fraction of synthetic tracks compared to all
identified tracks.  When there are no false detections the accuracy of
track creation is nearly 100\% because even with a full density solar
system model (for \ps-4) the sky-plane density of tracklets is low
enough to make confusion unimportant.  Since the next step in the MOPS
after track creation is to attempt an initial orbit determination
(IOD) on each identified track, the accuracy needs to be high in order
to not waste too many CPU cycles on attempting orbits on tracks that
are not valid.  However, calculating an IOD for tracks is trivially
parallelizable.

Note that the accuracy increases in table
\ref{tab.TrackIdentificationPerformance} in the first three steps of
increasing asteroid sky-plane density but drops precipitously on the
last jump.  This is due to the fact that we used a constant false
detection rate equal to the expected density of false detections in
all four simulations.  Thus, in the first three runs the noise is
dominated by the false detections but in the last run the density of
synthetic detections becomes high enough to add extra confusion into
the linking process.

At this point we have not put much effort into increasing the accuracy
of the \verb#linkTracklets# algorithm but there are many opportunities
to do so.  One such possibility is a multiple pass scenario in which
we first attempt to link the `easy' tracklets (\ie\ everything from
the Main Belt outwards) with relatively tight constraints on their
night-to-night motion, remove the tracklets that survive orbit
determination in good tracks, and then loosen the constraints in order
to identify difficult objects (NEOs).  We tested this technique on a
simulation involving over 22,000 tracklets in over 352,000 tracks.
Removing the properly linked tracklets, and all false tracks containing
any of those tracklets, left only about 6,700 tracks.  Thus, this
could be a powerful technique for increasing the effectiveness of the
inter-night linking process.

The decrease in accuracy of \verb#linkTracklets# at full density is
accompanied by a dramatic increase in the run time.  Increasing the
sky-plane density by a factor of about six increases the runtime by a
factor of almost thirty.  This is also not of particular concern
because the linking algorithm is easily parallelizable.  The
parallelization of \verb#linkTracklets# is easily implemented by
running each `pass' (described above) on a different processor.  The
sky-plane density of tracklets becomes high enough in the final
simulation of table \ref{tab.TrackIdentificationPerformance} to
require tripling the number of passes.

Tables \ref{tab.SSM250Tracks} through \ref{tab.SSMTracks} show the
progression of linking efficiency as a function of both the sky-plane
density and the solar system object type.  The efficiency is high for
all classes of objects and for all densities as would be expected
after table \ref{tab.TrackIdentificationPerformance}.  There is a very
slight decrease in linking efficiency for each object class as their
sky-plane density increases.  Within each model (each table) there is
a slight increase in linking efficiency with increasing mean
heliocentric distance of the object class.

The high efficiency for NEOs and distant objects should not be
surprising.  While their sky-plane density is low compared to the MB
objects, their rates of motion are often anomalous.  Even though the
distant objects do not move very far in a transient time interval and
therefore provide little motion vector information, the sky-plane
density of slow moving objects is low enough to make the linking
efficiency very high.

In \S\ref{s.SurveySimulation} it was pointed out that the survey
pattern avoids the 'stationary spots' and thus it must be remembered
that the results quoted herein do not provide results for all-sky
inter-night linking efficiency.  The intra-night linking efficiency
should be much better in the stationary spots because the detections
will be much closer together than at other points along the objects
orbits.  However, it is reasonable to expect that the inter-night
linking efficiency will decrease in the stationary spots due to the
unusual apparent acceleration of the objects in this region.

As mentioned earlier in this section, the MOPS has restricted its
requirements to linking only those objects with tracklets on three
nights within a lunation.  The more difficult problem of linking and
confirming just two nights of tracklets in one lunation or linking
three tracklets across two lunations is not handled with the
algorithms described here.  To extend the discovery phase space into
this realm we have teamed with Andrea Milani who will provide us
software capable of making these links.  The theoretical framework for
his work has been described elsewhere \citep{Mil05} and the realized
efficiencies in Pan-STARRS simulations will be discussed in a future
paper.




\subsection{MOPS and other surveys}
\label{ss.MOPSandOtherSurveys}

Contemporary wide-field asteroid surveys only perform the intra-night
linking step.  They identify asteroids by their linear motion in a
single night on three or more images.  The intra-night linking
efficiency has been measured by some of the major NEO surveys by
attempting to identify known asteroids in their fields.  The measured
peak efficiency for asteroids well above the limiting magnitude varies
from about 65\% (Spacewatch; \citet{Jed97}), $\sim 70$\% (Catalina Sky
Survey, observatory 703; Beshore personal communication), $\sim 90$\%
(Catalina Sky Survey, observatory G96; Beshore personal
communication), and about 90\% for the latest and reprocessed
Spacewatch data \citep{Lar01}.  In both these survey's the intra-night
linkings proposed by their algorithms are checked by a human observer.
This is clearly an impossible task at the Pan-STARRS discovery rate.

Some of the targeted (pencil-beam or narrow field) surveys have
determined their intra-night detection efficiency by injecting
synthetic asteroid images directly into the images before running
their source detection and linking algorithms \citep[\eg][]{Gla06,Pet04}.
They realize efficiencies of $\sim$90\%.

Inter-night linking is mostly performed by the Minor Planet Center
and there has been no report on their efficiency for this process.

To test the MOPS on real data before the onset of Pan-STARRS we have
obtained raw source detection lists from the Spacewatch
\citep{Lar01,Jed97} asteroid survey.  We have passed their data
through the MOPS and have identified apparently realistic asteroids.
In order to reduce the number of clearly false orbits identified by
MOPS we needed to run two pre-filters on the set of detections they
provide.  The first eliminates regions on the sky with unusual
over-densities of detections.  The over-densities are a problem in the
Spacewatch automated reduction process due to mis-estimating the
background level.  The second pre-filter reduces the prevalence of
anomalous sets of detections in linear features.  The Spacewatch
source finding algorithm identifies many false detections in the
linear features associated with bright star diffraction spikes, CCD
edge effects and artifical satellite streaks.

\begin{figure}
\epsscale{1.0}
\plotone{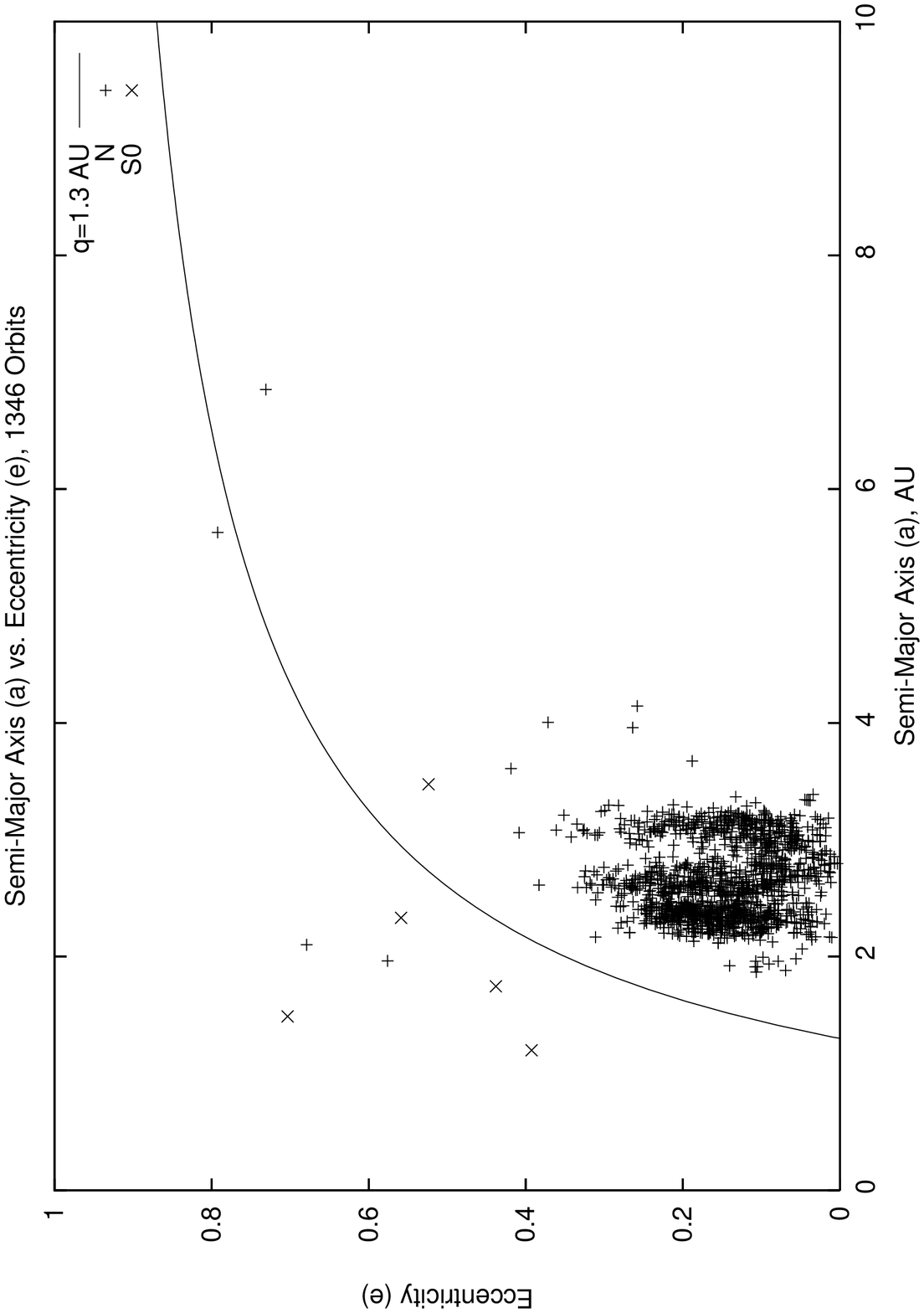}
\caption{Eccentricity versus semi-major axis for objects identified in
raw Spacewatch detections \ie in their entire transient source
detection list, not just the list of observations reported to the MPC.
The crosses represent presumably real objects identified in their
astrometry while the x's represent synthetic NEOs emplaced in the data
and subsequently identified by MOPS.  Note that one of the synthetic
NEOs is identified with a non-NEO orbit.}
\label{fig.Spacewatch.e.vs.a}
\end{figure}

Figure \ref{fig.Spacewatch.e.vs.a} shows the distribution of `derived'
objects - those objects for which MOPS formed tracklets, tracks,
initial orbit determination and differentially corrected orbits.
Since the figure shows final orbital elements for the derived objects it
goes beyond the purview of merely intra and inter-night linking as
discussed in the rest of this work. This is done for two reasons: 1)
because most of the Spacewatch detections are previously unknown
objects it would be difficult for us to establish which tracklets and
tracks were false and real and 2) to show that the MOPS is operational
on real data. The system efficiency through initial and differential
orbit determination will be described in a future paper.

\section {Conclusion}
\label{s.Conclusion}

The Pan-STARRS project has developed the first integrated asteroid
detection, intra and inter-night linking, attribution, precovery,
orbit identification and orbit determination system in the world.  It
is known as the Moving Object Processing System (MOPS).  For testing
and monitoring purposes during operations we have developed a
peudo-realistic simulation of the system including a realistic survey
strategy incorporating simple weather factors, S/N-dependent
astrometric noise and false detections at a sky-plane density expected
for the four telescope Pan-STARRS system.  The simulation does not
include additional important factors such as the camera fill factor or
probabilistic detections near the detection threshold.

We have developed new algorithms based on kd-tree and variable-trees
to link detections within and between nights that dramatically improve
the speed of identification and that scale as $O(\rho\log \rho)$ where
$\rho$ is the sky-plane density of objects.  The implementation of the
algorithms is trivially parallelizable on a set of CPU nodes.

Using these algorithms we have demonstrated nearly 100\% efficiency
for intra-night linking of synthetic detections with realistic
properties into `tracklets'.  Furthermore, we have demonstrated the
ability to obtain nearly 100\% efficiency for linking those tracklets
over many nights into `tracks'.  The accuracy of the algorithm, the
fraction of identified tracks that are actually synthetic in the
presence of noise, depends on and decreases with the sky-plane density
of detections.

Tests of the MOPS intra and inter-night linking algorithms on real
data provided by the Spacewatch facility show that the system is
capable of handling real data with all its inherent systematic
problems that are otherwise not explored in our synthetic surveying
model.


\clearpage

\section*{Acknowledgements}

The design and construction of the Panoramic Survey Telescope and
Rapid Response System by the University of Hawaii Institute for
Astronomy is funded by the United States Air Force Research Laboratory
(AFRL, Albuquerque, NM) through grant number F29601-02-1-0268.  The
MOPS is currently being developed in association with the Large
Synoptic Survey Telescope (LSST).  Jeremy Kubica's work was funded in
part by the LSST and by a grant from the Fannie and John Hertz
Foundation.  The LSST's research and development effort is funded in
part by the National Science Foundation under Scientific Program Order
No. 9 (AST-0551161) through Cooperative Agreement AST-0132798.
Additional funding comes from private donations, in-kind support at
Department of Energy laboratories and other LSSTC Institutional
Members.  Spacewatch (Robert McMillan) provided source detections for
testing the MOPS procedures.  Paulo Holvorcem provided timely
assistance for creating synthetic surveys.



\clearpage
\begin{deluxetable}{lcccccc}
\scriptsize
\tablewidth{0pc}
\tablecaption{\label{tab.StandardTrackletPerformance} 
  Standard tracklet identification performance in two regions}
\tablehead{
\colhead{Model} &
\colhead{Available} &
\colhead{Efficiency} &
\colhead{Accuracy} &
\colhead{Discordant} &
\colhead{Mixed} &
\colhead{Spurious}
}
\startdata
SSM Opposition  & 636251 & 99.97\% & 89.2\% & 3.8\% & 3.5\% & 3.5\% \\
SSM Sweet-Spots & 697927 & 99.97\% & 84.7\% & 8.7\% & 4.8\% & 1.9\% \\
\enddata

\tablecomments{Standard MOPS tracklet identification performance in
  the opposition and sweet spot regions for the full (\ps-4) solar system
  model (SSM) with full density false detections in a single lunation.
  Columns are: {\bf Available} - The number of possible synthetic
  tracklets that could be identified with detections separated by less
  than one hour; {\bf Efficiency} - the percentage of synthetic
  tracklets that were actually identified; {\bf Accuracy} - the
  percentage of all identified tracklets that were properly identified
  as being synthetic; {\bf Discordant} - the percentage of identified
  tracklets consisting of synthetic detections from different objects;
  {\bf Mixed} - the percentage of identified tracklets consisting of
  both synthetic and false detections; {\bf Spurious} - the percentage of
  identified tracklets consisting of false detections.}

\end{deluxetable}

\clearpage
\begin{deluxetable}{lcccccc}
\scriptsize
\tablewidth{0pc}
\tablecaption{\label{tab.NonStandardTrackletPerformance} 
  Non-Standard tracklet identification performance in two regions}
\tablehead{
\colhead{Model} &
\colhead{Available} &
\colhead{Efficiency} &
\colhead{Accuracy} &
\colhead{Discordant} &
\colhead{Mixed} &
\colhead{Spurious}
}
\startdata
SSM Opposition  & 636251 & 99.91\% & 30.7\% & 14.9\% & 27.8\% & 26.6\% \\
SSM Sweet-Spots & 698110 & 99.96\% & 26.4\% & 26.9\% & 34.0\% & 12.6\% \\
\enddata

\tablecomments{As in Table \ref{tab.StandardTrackletPerformance}, but
  for a non-standard MOPS implementation that ignores trailing
  information (orientation and length) for each detection.  i.e. each
  detection as treated as a simple point.}

\end{deluxetable}

\clearpage
\begin{deluxetable}{cccccccccccccc}
\rotate
\scriptsize
\setlength{\tabcolsep}{0.02in}
\tablewidth{0pc}
\tablecaption{\label{tab.TrackIdentificationPerformance} Overall Track Identification Performance}
\tablehead{
\colhead{Model} &	
\colhead{Objects} &	
\colhead{Density (\%)} &	
\colhead{Available} &	
\colhead{Linked} &	
\colhead{Efficiency (\%)} &
\colhead{Tracks} &	
\colhead{Accuracy} &	
\colhead{Overhead} &	
\colhead{Passes} &	
\colhead{Runtime (s)} &
\colhead{Rate (s$^{-1}$)} 													
}
\startdata
SSM/250 &     43445 &   0.4 &	680 &	   679 & 99.9 & 94041 &
  0.7 & 138.5 	& 112 	& 342 	& 127 \\					
MB/100&    960758 &   8.8 &	7658 &	  7644 & 99.8 & 138646 &
  5.5 & 18.1 	& 112 	& 387 	& 2483  \\
MB/10 &   1860758 &  17.1 &	21828 &	 21766 & 99.7 & 295529 &
  7.4 & 13.6 	& 112 	& 465 	& 4002 \\
SSM  &  10860758 & 100.0 &	156693 & 154109&  98.4  & 44814287 &
  0.3  & 290.8  & 361  	& 13642  & 796  \\
\enddata

\tablecomments{MOPS track identification performance for different
  solar system models (SSM).  The SSM is the full density (\ps-4)
  model and SSM/250 is every 250$^{th}$ object.  The MB/N models have
  full (\ps-4) densities of all SSM components except for the MB that
  includes every N$^{th}$ object.  Each model contains false
  detections at the full density level.  Columns are: {\bf Objects} -
  the number of different synthetic solar system objects included in
  the simulation; {\bf Density} - density of objects in the model
  compared to the full model; {\bf Available} - the number of
  synthetic tracks generated in the simulation; {\bf Linked} - the
  number of synthetic tracks that were properly linked; {\bf
  Efficiency} - the fraction of generated tracks that were correctly
  linked; {\bf Tracks} - the total number of tracks found in the
  simulation; {\bf Accuracy} - the percentage of identified tracks
  that represent synthetic tracks; {\bf Overhead} - the reciprocal of
  accuracy, the ratio of false to real tracks; {\bf Passes} -
  explained in the text in \S\ref{ss.IntraNightLinking}; {\bf Runtime}
  - on a single 3 GHz Pentium processor; {\bf Rate} - number of
  objects processed per second.}

\end{deluxetable}

\clearpage
\begin{deluxetable}{crrr}
\scriptsize
\tablewidth{0pc}
\tablecaption{\label{tab.SSM250Tracks} SSM/250 Track identification
  performance by object type}
\tablehead{
\colhead{Object Type} &
\colhead{Available} &
\colhead{Linked} &
\colhead{Efficiency}
}
\startdata
NEO  	 &	2  &	2  &	100.0\% \\
MB 	&	656 &	655 &	99.8\% \\ 
TRO 	&	18 &	18 &	100.0\% \\ 
CEN 	&	0 &	0 &	N/A 	\\
TRO 	&	4 &	4 &	100.0\% \\ 
TNO 	&	0 &	0 &	N/A 	\\
COM 	&	0 &	0 &	N/A 	\\
\hline
Total    &       680 &	679 &	99.9\% \\ 														
\enddata

\tablecomments{MOPS track identification performance in a single
  lunation by solar system object type for a solar system model with
  only every 250$^{th}$ object.  Columns are: {\bf Object Type} - the
  class of solar system object with obvious abbreviations; {\bf
  Available} - the number of synthetic tracks generated in the
  simulation; {\bf Linked} - the number of synthetic tracks that were
  properly linked; {\bf Efficiency} - the fraction of generated tracks
  that were correctly linked.}

\end{deluxetable}

\clearpage
\begin{deluxetable}{crrr}
\scriptsize
\tablewidth{0pc}
\tablecaption{\label{tab.MB100Tracks} MB/100 Track identification
  performance by object type}
\tablehead{
\colhead{Object Type} &
\colhead{Available} &
\colhead{Linked} &
\colhead{Efficiency}
}
\startdata
NEO  	 &	351  &	342  &	97.4\% \\
MB 	&	1626 &	1624 &	99.9\% \\
TRO 	&	4425 &	4423 &	100.0\% \\
CEN 	&	99 &	99  &	100.0\%  \\
TNO	&	818 &	818 &	100.0\%   \\
TNO	&	307 &	307 &	100.0\%  \\
COM	&	32 &	31 &	96.9\%  \\
\hline
Total           &     7658 &	7644 &	99.8\% 	 \\								
\enddata

\tablecomments{MOPS track identification performance by solar system
  object type for a solar system model with only every 100$^{th}$ main
  belt object and all other objects at full (\ps-4) density.  Columns
  are as in Table \ref{tab.SSM250Tracks}.}

\end{deluxetable}

\clearpage
\begin{deluxetable}{crrr}
\scriptsize
\tablewidth{0pc}
\tablecaption{\label{tab.MB10Tracks} MB/10 Track identification
  performance by object type}
\tablehead{
\colhead{Object Type} &
\colhead{Available} &
\colhead{Linked} &
\colhead{Efficiency}
}
\startdata
 NEO  	 &	351  &	343  &	97.7\% \\
MB 	&	15769 &	15723 &	99.7\% \\
TRO 	&	4465 &	4460 &	99.9\% \\ 
CEN 	&	106 &	105 &	99.1\% \\ 	
TRO 	&	818 &	818 &	100.0\% \\ 
TNO 	&	287 &	286 &	99.7\% \\ 
COM 	&	32 &	31 &	96.9\% \\ 	
\hline
Total  & 21828 &	21766 &	99.7\% \\ 		
\enddata

\tablecomments{MOPS track identification performance by solar system
  object type for a solar system model with only every 10$^{th}$ main
  belt object and all other objects at full (\ps-4) density.  Columns
  are as in Table \ref{tab.SSM250Tracks}.}
\end{deluxetable}

\clearpage
\begin{deluxetable}{crrr}
\scriptsize
\tablewidth{0pc}
\tablecaption{\label{tab.SSMTracks} Full SSM Track identification
  performance by object type}
\tablehead{
\colhead{Object Type} &
\colhead{Available} &
\colhead{Linked} &
\colhead{Efficiency}
}
\startdata
 NEO  	 &	350  &	340  &	97.1\% \\
MB 	&	151084 &	148526 &	98.3\% \\ 
TRO 	&	4161 &	4148 &	99.7\% \\ 	
CEN 	&	99 &	99 &	100.0\% \\ 
TRO 	&	695 &	693 &	99.7\% \\ 
TNO 	&	275 &	274 &	99.6\% \\ 	
COM 	&	29 &	29 &	100.0\% \\ 
Total    &    156693 & 	154109 &	98.4\% \\ 			
\enddata

\tablecomments{MOPS track identification performance by solar system
  object type for the full (\ps-4) density solar system model.
  Columns are as in Table \ref{tab.SSM250Tracks}.}

\end{deluxetable}



\clearpage

\clearpage

\clearpage

\clearpage

\clearpage
\begin{figure} [!tb]
\begin{center}
\begin{tabular}{ll}
\hline
& \textbf{Recursive kd-Tree Range Search} \\
& \textbf{Input}: Current Tree Node ${\mathbf T}$, query point ${\mathbf q}$, radius $r$. \\
& \textbf{Output}: A list of matching points ${\mathbf Z}$ \\
\hline
1. & IF ${\mathbf q}$ is within $r$ node ${\mathbf T}$'s bounding box: \\
2. & \hspace{0.20in} IF ${\mathbf T}$ is a leaf node: \\
3. & \hspace{0.40in} FOR each data point ${\mathbf x}$ owned by node ${\mathbf T}$: \\
4. & \hspace{0.60in} IF ${\mathbf q}$ is within $r$ of ${\mathbf x}$: \\
5. & \hspace{0.80in} Add ${\mathbf x}$ to ${\mathbf Z}$. \\
6. & \hspace{0.20in} ELSE \\
7. & \hspace{0.40in} Recursively search using each ${\mathbf T}$'s children nodes in place of ${\mathbf T}$. \\
8. & Return ${\mathbf Z}$ \\
\hline
\end{tabular}
\caption{A recursive search for points within radius $r$ of the query
point ${\mathbf q}$ using a kd-tree.  This search is initially called
with the root node of the kd-tree.} \label{fig:kdrangesearch}
\end{center}
\end{figure}

\clearpage

\clearpage
\begin{figure} [!tb]
\begin{center}
\begin{tabular}{ll}
\hline
& \textbf{Sequential Intra-Night Linkage Algorithm} \\
& \textbf{Input}: A set ${\mathbf X}$ of all input detections \\
& \textbf{Output}: A list of result tracks ${\mathbf Z}$ \\
\hline
1.   & Build a kd-Tree on the detections ${\mathbf X}$.\\
2.   & FOR each tracklet ${\mathbf x} \in {\mathbf X}$: \\
3.   & \hspace{0.20in} Use the kd-tree to efficiently find ${\mathbf Y}$ the set of all reachable detections. \\
4.   & \hspace{0.20in} FOR each potential pairing $({\mathbf x} \in {\mathbf X}, {\mathbf y} \in {\mathbf Y})$: \\
5.   & \hspace{0.40in} Create a new linear tracklet ${\mathbf z}$ from ${\mathbf x}$ and ${\mathbf y}$.\\
6.   & \hspace{0.40in} Use the kd-tree (or the set ${\mathbf Y}$) to find all supporting detections \\
& \hspace{0.40in} compatible with ${\mathbf z}$. \\
7.   & \hspace{0.40in} IF ${\mathbf z}$ has enough support. \\
8.   & \hspace{0.60in} Add ${\mathbf z}$ to ${\mathbf Z}$. \\
9. & Return ${\mathbf Z}$ \\
\hline
\end{tabular}
\caption{A simplified multiple hypothesis tracking algorithm for
asteroid linkage.}\label{fig:sequentialalg}
\end{center}
\end{figure}

\clearpage

\clearpage
\begin{figure} [!tb]
\begin{center}
\begin{tabular}{ll}
\hline
& \textbf{Moving Object Range Search} \\
& \textbf{Input}: A query point ${\mathbf q}$, a current tree node ${\mathbf T}$, and \\
& \hspace{0.4in} minimum and maximum speeds: $v_{min}$ and $v_{max}$. \\
& \textbf{Output}: A list of feasible points ${\mathbf Z}$ \\
\hline
1. & If we cannot prune ${\mathbf T}$ as per Equations \ref{eq:movingrangeprune1} and \ref{eq:movingrangeprune2}: \\
2. & \hspace{0.15in} IF ${\mathbf T}$ is a leaf node: \\
3. & \hspace{0.30in} FOR each ${\mathbf x}$ owned by ${\mathbf T}$: \\
4. & \hspace{0.45in} $v = \frac{dist({\mathbf q}, {\mathbf x})}{| t_q - t_x |}$ \\
5. & \hspace{0.45in} IF $v_{min} \leq v \leq v_{max}$: \\
6. & \hspace{0.60in} Add ${\mathbf x}$ to ${\mathbf Z}$. \\
7. & \hspace{0.15in} ELSE: \\
8. &\hspace{0.30in} Recursively search using ${\mathbf T}$'s left child. \\
9. &\hspace{0.30in} Recursively search using ${\mathbf T}$'s right child. \\
10. & Return ${\mathbf Z}$. \\
\hline
\end{tabular}
\caption{The recursive algorithm for a moving object range search.}
\label{fig:movingobjsearchalg}
\end{center}
\end{figure}

\clearpage

\clearpage

\clearpage
\begin{figure} [!tb]
\begin{center}
\begin{tabular}{ll}
\hline
& \textbf{Variable-Tree Model Detection} \\
& \textbf{Input}: A set of $M$ current model tree nodes ${\mathbf M}$ \\
& \hspace{0.35in} A set of current support tree nodes ${\mathbf S}$ \\
& \textbf{Output}: A list ${\mathbf Z}$ of feasible sets of points \\
\hline
1. & ${\mathbf S}' \gets \{ \}$ and ${\mathbf S}_{curr} \gets {\mathbf S}$ \\
2. & IF we cannot prune based on the mutual compatibility of ${\mathbf M}$: \\
3. & \hspace{0.15in} FOR each ${\mathbf s} \in {\mathbf S}_{curr}$ \\
4. & \hspace{0.3in} IF ${\mathbf s}$ is compatible with ${\mathbf M}$: \\
5. & \hspace{0.45in} IF ${\mathbf s}$ is ``too wide'': \\
6. & \hspace{0.60in} Add ${\mathbf s}$'s left and right child to the end of ${\mathbf S}_{curr}$. \\
7. & \hspace{0.45in} ELSE \\
8. & \hspace{0.60in} Add ${\mathbf s}$ to ${\mathbf S}'$. \\
9. & \hspace{0.15in} IF we have enough valid support points: \\
10. & \hspace{0.30in} IF all of ${\mathbf m} \in {\mathbf M}$ are leaves: \\
11. & \hspace{0.45in} Test all combinations of points owned by the model nodes, using \\
& \hspace{0.50in}  the support nodes' points as potential support.  \\
 & \hspace{0.45in} Add valid sets to ${\mathbf Z}$. \\
12. & \hspace{0.30in} ELSE \\
13.& \hspace{0.45in} Let ${\mathbf m}^*$ be the non-leaf model tree node that owns the most points.  \\
14. &\hspace{0.45in} Search using ${\mathbf m}^*$'s left child in place of ${\mathbf m}^*$ and ${\mathbf S}'$ instead of ${\mathbf S}$. \\
15. &\hspace{0.45in} Search using ${\mathbf m}^*$'s right child in place of ${\mathbf m}^*$ and ${\mathbf S}'$ instead of ${\mathbf S}$. \\
\hline
\end{tabular}
\caption{A simple variable-tree algorithm for spatial structure
search.  This algorithm uses simple heuristics such as:
searching the model node with the most points and splitting a support
node if it is too wide.  These heuristics can be replaced by more
accurate, problem-specific ones.} \label{fig:multtreealg}
\end{center}
\end{figure}

\clearpage

\clearpage


\clearpage

\begin{thebibliography}{widest-label}

\bibitem[Bar-Shalom \& Li (1995)]{Bar95}
Bar-Shalom,~Y., and Li,~X-R.\ 1995.\ Multitarget-Multisensor Tracking:
Principles and Techniques.\ Published by Yaakov Bar-Shalom, Storrs,
CT.

\bibitem[Bar-Shalom \etal (2001)]{Bar01}
Bar-Shalom,~Y., Li,~X-R, and Kirubarajan, T.\ 2001.\ Estimation with
Applications to Tracking and Navigation: Theory Algorithms and
Software.\ Published by Wiley-Interscience, New York, NY.

\bibitem[Bentley (1975)]{Ben75} Bentley, J.~L.\ 1975.\
Multidimensional Binary Search Trees Used for Associative Searching.
Communications of the ACM, 18, 9, 509-517.

\bibitem[Bernstein \etal (2004)]{Ber04} 
Bernstein, G.~M., Trilling, D.~E., Allen, R.~L., Brown, M.~E., Holman,
M., Malhotra, R.\ 2004.\ The Size Distribution of Trans-Neptunian
Bodies.\aj128, 1364-1390.

\bibitem[Blackman \& Popoli (1999)]{Bla99}
Blackman,~S. and Popoli,~R.\ 1999.\ Design and Analysis of Modern
Tracking Systems.  Published by Artech House, Boston, MA.

\bibitem[Boattini \etal (2006)]{Boa06}
Boattini,~A. and 11 co-authors.\ 2006.\ 
The Campo Imperatore Near Earth Object Survey (CINEOS).\
Submitted to `Earth, Moon and Planets'.

\bibitem[Bottke \etal (2002)]{Bot02} 
Bottke, W.~F., Morbidelli, A., Jedicke, R., Petit, J.-M., Levison,
H.~F., Michel, P., and Metcalfe, T.~S.\ 2002.\ Debiased Orbital and
Absolute Magnitude Distribution of the Near-Earth Objects.\ Icarus
156, 399-433.

\bibitem[Bowell \etal (1994)]{Bow94a}
E.~Bowell, K.~Muinonen, and L.~H.~Wasserman.\ 1994.\
 A public-domain asteroid orbit database. 
In "Asteroids, Comets, Meteors 1993" (A. Milani et al., eds.),
pp. 477-481. Kluwer, Dordrecht

\bibitem[Bowell \& Muinonen(1994)]{Bow94b} 
Bowell, E.~and Muinonen, K.\ 1994.\  
Earth-crossing Asteroids and Comets: Groundbased Search Strategies.\  
Hazards Due to Comets and Asteroids, p. 149. 

\bibitem[Bowell \etal (1989)]{Bow89} 
Bowell, E., Hapke, B., Domingue, D., Lumme, K., Peltoniemi, J.,
Harris, A.~W.\ 1989.\ Application of photometric models to asteroids.\
Asteroids II 524-556.

\bibitem[Chesley \& Spahr (2004)]{Che04} 
Chesley, S.~R.~and Spahr, T.~B.\ 2004.\ Earth impactors: orbital
characteristics and warning times.\ Mitigation of Hazardous Comets and
Asteroids, 22.

\bibitem[Denneau \etal (2006a)]{Den06a}
Denneau,~L., .\ 2006.\  The Pan-STARRS Solar System Model. In preparation.

\bibitem[Denneau \etal (2006b)]{Den06b} 
Denneau,~L.\ 2006.\ The Pan-STARRS Solar System Survey Simulation.
In preparation.

\bibitem[Elliot \etal (2005)]{Ell05} 
Elliot, J.~L., and 10 colleagues 2005.\ The Deep Ecliptic Survey: A
Search for Kuiper Belt Objects and Centaurs. II. Dynamical
Classification, the Kuiper Belt Plane, and the Core Population.\aj129,
1117-1162.

\bibitem[Gladman \etal (2006)]{Gla06}
Gladman, B. J. and 11 co-authors.\ 2006.\ 
A sub-kilometer asteroid diameter survey.\ 
Submitted to \icarus.

\bibitem[Granvik \& Muinonen (2005)]{Gra05} 
Granvik, M.~and Muinonen, K.\ 2005.\ 
Asteroid identification atdiscovery.\ 
Icarus 179, 109-127.

\bibitem[Gauss(1809)]{Gau1809} 
Gauss, K.~F.\ 1809.\
Hambvrgi, Svmtibvs F.~Perthes et I.~H.~Besser, 1809.

\bibitem[Harris (1998)]{Har98} 
Harris, A.~W.\ 1998.\ Evaluation of ground-based optical surveys for
near-Earth asteroids.\ \planss 46, 283-290.

\bibitem[Hodapp \etal (2004)]{Hod04} 
Hodapp, K.~W., and 30 colleagues 2004.\ Design of the Pan-STARRS
telescopes.\ Astronomische Nachrichten 325, 636-642.

\bibitem[Jedicke \etal (2003)]{Jed03} 
Jedicke, R., Morbidelli, A., Spahr, T., Petit, J.-M., and Bottke,
W.~F.\ 2003.\ Earth and space-based NEO survey simulations: prospects
for achieving the spaceguard goal.\ Icarus 161, 17-33.

\bibitem[Jedicke \etal(2002)]{Jed02} 
Jedicke, R., Larsen, J., and Spahr, T.\ 2002.\ Observational Selection
Effects in Asteroid Surveys.\ Asteroids III, 71-87.

\bibitem[Jedicke \& Herron (1997)]{Jed97} 
Jedicke, R.~and Herron, J.~D.\ 1997.\ Observational Constraints on the
Centaur Population.\ Icarus 127, 494-507.

\bibitem[Jedicke (1996)]{Jed96} 
Jedicke, R.\ 1996.\ Detection of Near Earth Asteroids Based Upon Their
Rates of Motion.\ \aj 111, 970.

\bibitem[Jewitt \etal (2000)]{Jew00} 
Jewitt, D.~C., Trujillo, C.~A., Luu, J.~X.\ 2000.\ Population and Size
Distribution of Small Jovian Trojan Asteroids.\aj120, 1140-1147.


\bibitem[Kaiser (2004)]{Kai04}
Kaiser,~N.\  2004.\ .  The Likelihood of Point Sources in Pixellated
Images.\ Pan-STARRS internal document PSDC-200-010-00
available upon request.

\bibitem[Kristensen (2004)]{Kri04} 
Kristensen, L.~K.\ 2004.\ Initial Orbit Determination for Distant
Objects.\ \aj127, 2424-2435.

\bibitem[Kristensen (2002)]{Kri02} 
Kristensen, L.~K.\ 2002.\ Follow-up Ephemerides and the Accuracy of
Preliminary Orbits.\ Icarus 159, 339-350.

\bibitem[Kristensen (1992)]{Kri92} 
Kristensen, L.~K.\ 1992.\ The identification problem in asteroid
surveys.\ \aap 262, 606-612.

\bibitem[Kubica \etal (2005a)]{Kub05a}
Kubica,~J., Moore,~A., Connolly, R., and Jedicke,~R.\ 2005.\ A
Multiple Tree Algorithm for the Efficient Association of Asteroid
Observations.\ The Eleventh ACM SIGKDD International Conference on
Knowledge Discovery and Data Mining (2005), ACM Press, Eds. Robert
L. Grossman and Roberto Bayardo and Kristin Bennett and Jaideep Vaidya,
p. 138-146.

\bibitem[Kubica \etal (2005b)]{Kub05b}
Kubica,~J., Masiero,~J., Moore,~A., Jedicke,~R., and Connolly, R.\
2005.\ Variable kd-Tree Algorithms for Spatial Pattern Search and
Discovery.\ Advances in Neural Information Processing Systems, MIT
Press, Eds. Y. Weiss and B. Sch\"{o}lkopf and J. Platt, p. 691-698.

\bibitem[Larsen \etal (2001)]{Lar01} 
Larsen, J.~A., Gleason, A.~E., Danzl, N.~M., Descour, A.~S., McMillan,
R.~S., Gehrels, T., Jedicke, R., Montani, J.~L., Scotti, J.~V.\ 2001.\
The Spacewatch Wide-Area Survey for Bright Centaurs and
Trans-Neptunian Objects.\ 
Astronomical Journal 121, 562-579.

\bibitem[Masiero \etal (2006)]{Mas06}
Masiero,~J. \etal\ 2006.\  The CFHT Main Belt G-survey.  in preparation.

\bibitem[Marsden(1985)]{Mar85} 
Marsden, B.~G.\ 1985.\ Initial orbit determination - The pragmatist's
point of view.\ \aj 90, 1541-1547.

\bibitem[McKee \& Taylor \etal (2001)]{McK01} 
McKee, C. F. with Taylor, T. F. and 13 co-authors.\ 2001.\ Astronomy
and Astrophysics in the New Millennium.\ Library of Congress Card
Number: 00-112257, National Academy Press, Washington, DC.

\bibitem[Mignard (2002)]{Mig02} 
Mignard, F.\ 2002.\ Observations of solar system objects with
GAIA. I. Detection of NEOS.\ \aap 393, 727-731.

\bibitem[Milani \etal (2006)]{Mil06} 
Milani, A., Gronchi, G.~F., Kne{\v z}evic, Z., Sansaturio, M.~E.,
Arratia, O., Denneau, L., Grav, T., Heasley, J., Jedicke, R., Kubica,
J.\ 2006.\ Unbiased orbit determination for the next generation
asteroid/comet surveys.\ IAU Symposium 229, 367-380.

\bibitem[Milani \etal (2005)]{Mil05} 
Milani, A., Gronchi, G.~F., Kne{\v z}evi{\'c}, Z., Sansaturio, M.~E.,
and Arratia, O.\ 2005.\ Orbit determination with very short arcs.\
Icarus 179, 350-374.

\bibitem[Milani \etal (2001)]{Mil01} 
Milani, A., Sansaturio, M.~E., and Chesley, S.~R.\ 2001.\ The Asteroid
Identification Problem IV: Attributions.\ Icarus 151, 150-159.

\bibitem[Milani \etal (2000)]{Mil00} 
Milani, A., La Spina, A., Sansaturio, M.~E., and Chesley, S.~R.\
2000.\ The Asteroid Identification Problem. III. Proposing
Identifications.\ Icarus 144, 39-53.

\bibitem[Milani \& Valsecchi (1999)]{Mil99b} 
Milani, A.~and Valsecchi, G.~B.\ 1999.\ The Asteroid Identification
Problem. II. Target Plane Confidence Boundaries.\ Icarus 140, 408-423.

\bibitem[Milani (1999)]{Mil99a} 
Milani, A.\ 1999.\ The Asteroid Identification Problem. I. Recovery of
Lost Asteroids.\ Icarus 137, 269-292.

\bibitem[Morrison (1992)]{Mor92} 
Morrison, D.\ 1992.\ The Spaceguard Survey - Protecting the earth from
cosmic impacts.\ Mercury 21, 103-106.

\bibitem[Petit \etal (2004)]{Pet04} 
Petit, J.-M., Holman, M., Scholl, H., Kavelaars, J., Gladman, B.\ 2004.\ 
A highly automated moving object detection package.\ 
\mnras 347, 471-480.

\bibitem[Raymond \etal (2004)]{Ray04} 
Raymond, S.~N., and 23 colleagues 2004.\ A Strategy for Finding
Near-Earth Objects with the SDSS Telescope.\ \aj 127, 2978-2987.

\bibitem[Stokes \etal (2002)]{Sto02} 
Stokes, G.~H., Evans, J.~B., and Larson, S.~M.\ 2002.\ Near-Earth
Asteroid Search Programs.\ Asteroids III, 45-54.

\bibitem[Tedesco \etal (2005)]{Ted05} 
Tedesco, E.~F., Cellino, A., and Zappal{\'a}, V.\ 2005.\ The
Statistical Asteroid Model. I. The Main-Belt Population for Diameters
Greater than 1 Kilometer.\ \aj 129, 2869-2886.

\bibitem[Tonry \etal (1997)]{Ton97} 
Tonry, J., Burke, B.~E., Schechter, P.~L.\ 1997.\ 
The Orthogonal Transfer CCD.\ 
\pasp 109, 1154-1164.

\bibitem[Yoshida et al.(2003)]{Yos03} 
Yoshida, F., Nakamura, T., Watanabe, J.-I., Kinoshita, D., Yamamoto,
N., Fuse, T.\ 2003.\ Size and Spatial Distributionsof Sub-km Main-Belt
Asteroids.\ \pasj 55, 701-715.

\bibitem[Virtanen \etal (2001)]{Vir01} 
Virtanen, J., Muinonen, K., and Bowell, E.\ 2001.\ Statistical Ranging
of Asteroid Orbits.\ Icarus 154, 412-431.

\end{thebibliography}
\end{document}